\documentclass[a4paper,prb,superscriptaddress, amsfonts, amssymb, amsmath, reprint, showkeys, nofootinbib, twoside]{revtex4-2}
\usepackage[english]{babel}
\usepackage[utf8]{inputenc}
\usepackage{amsthm}
\usepackage{mathtools}
\usepackage{physics}
\usepackage{xcolor}
\usepackage{graphicx}
\usepackage[T1]{fontenc}
\usepackage[pdftex, pdftitle={Article}, pdfauthor={Author}]{hyperref} 
\usepackage{physics}
\usepackage[colorinlistoftodos, color=green!40, prependcaption]{todonotes}
\usepackage{dsfont}
\usepackage{chemformula}
\usepackage{units}
\usepackage{amsmath}
\usepackage{relsize}
\usepackage{enumitem}
\usepackage{epsfig}
\usepackage[amssymb]{SIunits}
\usepackage[normalem]{ulem}

\begin{document}
\title{Spin waves in alloys at finite temperatures: application  for FeCo magnonic crystal} 
\author{Sebastian Paischer} \email{sebastian.paischer@jku.at} 
\affiliation{Institute for  Theoretical Physics, Johannes Kepler  University Linz, Altenberger  Stra{\ss}e 69, 4040 Linz} 
\author{Pawe\l{} A. Buczek}
\affiliation{Department of Engineering and Computer Sciences, Hamburg  University of Applied Sciences, Berliner Tor 7, 20099 Hamburg, Germany} 
\author{Nadine Buczek} 
\affiliation{Department of Applied Natural Sciences, L\"ubeck  University of Applied Sciences, M\"onkhofer Weg 239, 23562 L\"ubeck,  Germany} 
\author{David Eilmsteiner} 
\affiliation{Institute for  Theoretical Physics, Johannes Kepler  University Linz, Altenberger  Stra{\ss}e 69, 4040 Linz} 
\author{Arthur Ernst}
\affiliation{Max-Planck-Institut of Microstructure Physics, Weinberg  2, 06120 Halle (Saale), Germany} 
\affiliation{Institute for Theoretical Physics, Johannes Kepler University Linz, Altenberger  Stra{\ss}e 69, 4040 Linz} 
\date{\today}

\begin{abstract}
  We study theoretically the influence of the temperature and
  disorder on the spin wave spectrum of the magnonic crystal
  $\ch{Fe_{1-c}Co_{c}}$. Our formalism is based on the analysis of a
  Heisenberg Hamiltonian by means of the wave vector and
  frequency dependent transverse magnetic susceptibility. The exchange integrals entering the model are obtained from the \emph{ab initio} magnetic force theorem. The coherent potential approximation is employed to treat the disorder and random phase approximation in order to account for the softening of the magnon spectrum at finite temperatures. The alloy turns out to exhibit many advantageous properties for spintronic applications.
  Apart from high Curie temperature, its magnonic bandgap remains
  stable at elevated temperatures and is largely unaffected by the
  disorder.  We pay particular attention to the attenuation of magnons
  introduced by the alloying.  The damping turns out to be a
  non-monotonic function of the impurity concentration due to the
  non-trivial evolution of the value of exchange integrals with the Co
  concentration.  The disorder induced damping of magnons is estimated
  to be much smaller than their Landau damping.
\end{abstract}


\maketitle

\section{Introduction}

Magnon spintronics, or magnonics, is a novel promising strategy in the
engineering of data processors \cite{Chumak2015}.  It takes advantage
of spin waves (also called magnons) in order to perform logical
computations \cite{Khitun2011,AlWahsh2011}.  Magnons emerge as
collective excitations of magnetically ordered solids and can be
pictured as wave-like coherent precession of atomic moments
\cite{Bloch1930}.  In periodic structures, including atomic lattices,
these quasiparticles are Bloch waves, carrying energy and crystal
momentum.  Magnonic computers avoid numerous drawbacks of classical
semiconductor based computers, but they rely heavily on suitably
designed magnon propagation media.  Their particularly relevant class
are \textit{magnonic crystals} \cite{Chumak2017,Nikitov2001} featuring
spin wave propagation properties not typically found in common
magnetic solids like elemental ferro- and antiferromagnets, especially
the emergence of a \textit{magnonic gap}, i.e. frequency bands in which
magnon states cannot propagate in the solid
\cite{Krawczyk2014,Lenk2011}. This feature, combined with the unique
spin wave dispersion close to the band edges, provides a rich toolbox for magnon mode engineering, including the possibility of
selective spin wave excitations and propagation, magnon mode
confinement and deceleration, and bandgap soliton generation \cite{Sadovnikov2016,Sadovnikov2018,Sheshukova2013}.

The bulk of the current research in this domain revolves around the
utilization of long wave-length magnons with energies in the gigahertz
band.  Nevertheless, in order to definitely push the size and speed
limits of modern semiconductor computers, one must resort to the
spin waves in the terahertz regime.  While the foundations for the
magnonic information processing in the terahertz regime are laid, the
potential of the terahertz magnonics remains vastly unexplored
\cite{Zakeri2018}.  At the same time, one expects well defined
spin waves in this energy range \cite{Buczek2010d} and in systems
with many different atoms in the primitive cell, the modes may well
arrange in bands separated by the magnonic gap \cite{Buczek2009},
yielding natural magnonic crystals.

Here, we concentrate on the ferromagnetic \ch{Fe_{1-c}Co_{c}} alloy.
With typical magnon energies well within the terahertz range, a high
Curie temperature \cite{Normanton1975,Nishizawa1984} and the bandgap
in the spectrum,
opening due to the the large difference in the interaction
strengths and magnetic moments of the constituents and remaining stable at elevated
temperatures, the alloy family shows all the necessary properties for
a terahertz magnonic crystals.  It is interesting to note that the
magnonic crystals used in terahertz applications are typically
artificial heterostructures obtained from elaborate fabrication
processes.  On the contrary, in the terahertz range, the natural
microscopic arrangement of atoms in alloys like \ch{Fe_{1-c}Co_{c}}
would suffice to create cheap magnonic crystals.

In metals, the lifetime of the modes is limited by the
interaction of these collective modes with the single particle
continuum, called Landau damping
\cite{Buczek2011,Buczek2011a,Zhang2012}, but means of viable
engineering of long-living magnons have been proposed, such as
reducing the system's dimensionality and alloying \cite{Qin2015}.  The
latter method leads to a further momentum dissipation mechanism, in
which the Bloch waves cease to be the eigenstates of the magnetic
Heisenberg-like Hamiltonian and acquire a finite lifetime arising from
the scattering on the crystal imperfections
\cite{Buczek2016,Buczek2018}.  This picture of the weak attenuation
might break down if the magnon spectra become dominated by strongly
spatially localized modes. 
Further mechanism limiting the lifetime
of the magnon modes, and thus their potential to propagate
dissipationlessly through the medium, is the interaction of the modes
with a thermal bath.\\

Solids, and in particular nanostructures, often feature structural
imperfections.  Furthermore, in order to be usable, the magnonic
computers must be able to operate at and typically well above room
temperature.  Thus, it is prudent to delve into the central question
of this paper, namely how the magnonic properties evolve in real,
imperfect or alloyed solids at non-zero temperatures.  We show that
the increase of the disorder in \ch{Fe_{1-c}Co_{c}} alloy not only
preserves the magnonic gap but can even be used to precisely engineer
its value and further properties.

Among others, we address the influence of temperature and disorder
on the magnonic band gap as well as on the dispersion and lifetimes
of the spin waves.
Our formalism is based on the coherent potential approximation (CPA)
applied to the disordered Heisenberg ferromagnet \cite{Buczek2016}. This description of magnetic degrees of freedom with an effective Heisenberg Hamiltonian, although originally put forth for the description of magnetic insulators, has been shown to yield remarkably accurate results for metallic magnets also beyond the long wavelength limit in which it can be shown to be formally accurate \cite{Zakeri2014,Halilov_1998,Etz2015,Pajda2001}.
The superiority of our method compared to other treatments of the same
problem is the possibility to account for complex crystal
structures. 
To incorporate finite temperature effects, we
implemented a modified version of the random phase approximation
(RPA) discussed in reference \cite{Callen1963}. Our formalism does not
include the Landau damping of the spin waves. This attenuation
mechanism can be pronounced in metallic magnonic crystals and can be
described within the framework of many-body perturbation theory (MBT)
\cite{Bluegel2013,Okumura2019} or time-dependent density functional theory (TDDFT)
\cite{Buczek2011a} (called ``dynamical methods''). Unfortunately, at the moment, no
feasible formal and computational methodology allowing to incorporate
the effects of disorder into these two approaches has been put forward. However, those dynamical methods reveal that in most ordered 3d itinerant
magnets, except for moderate Landau broadening, well defined magnons are indeed expected
in the entire Brillouin zone, possibly with the exception of bcc Fe where
“spin-wave disappearance” phenomenon sets in in limited parts of the Brillouin zone \cite{Buczek2011a}. Relevant for the case of FeCo, the MBT calculations of {\c{S}}a{\c{s}}{\i}o{\u{g}}lu \emph{et. al.} \cite{Sasiouglu2013} predict
the existence of well defined spin-waves in the entire zone both for the acoustic and optical
modes. 
Correspondingly, neutron scattering experiments, cf. references in \cite{Buczek2011a}, and spin-polarized
electron energy loss spectroscopy \cite{Zhang2010,Etzkorn2005,Tang2007,Vollmer2003} probe clear high energy spin-wave signals in the
entire zone. Although neither MFT nor TDDFT can properly describe spin waves in Ni, magnons in Co and Fe are well reproduced for almost the whole BZ and a large frequency range (see e.g.\cite{Tang2007,Vollmer2003}). An even better agreement between the MFT calculations and experiment is achieved in layered materials \cite{Meng2014,Chuang2014,Zakeri2013,Qin2019,Zakeri2021,Qin2013}, caused by significantly weaker interaction with the Stoner continuum compared to 3D materials. When the associated Heisenberg model is
employed at elevated temperatures it yields very good account of the phase transition temperatures (even in bcc Fe), indicating that the high temperature phase is essentially correctly captured as well \cite{Rusz2005,Rusz2006}. Thus it is reasonable to apply the Heisenberg model to study the impact of disorder on the damping of spin waves.

The paper is organized as follows: In chapter \ref{chap_theory}, the
theoretical background of the RPA-CPA theory for the disordered
Heisenberg ferromagnet is discussed. Some numerical details are given
in section \ref{chap_implementation}. The results are
presented in chapter \ref{chap_res}.

\section{Theory}\label{chap_theory}
The Heisenberg ferromagnet is characterized by the Hamiltonian
\begin{align}
  H=-\frac{1}{2}\sum_{i,j}J_{ij}~\vb*{e}_i\cdot\vb*{e}_j
\end{align}
where $J_{ij}$ are the exchange parameters which can be obtained from
the magnetic force theorem \cite{Liechtenstein1987,Solovyev2021} and $\vb*{e}_i$ is
a unit vector in the direction of the magnetic moment. To calculate
magnon-properties, the transverse susceptibility \cite{Nolting2009}
\begin{align}
  \chi_{ij}(t,t')=-\text{i}~\Theta(t-t')~\overline{\comm{\mu_i^+(t)}{\mu_j^-(t')}}\label{eqn_susc_def}
\end{align}
with $\mu_i^\pm=\mu_i^x\pm\text{i}\mu_i^y$, $\mu_i^\alpha$ being the
$\alpha$-component of the magnetic moment $\vb*{\mu}_i$ on the lattice
site $i$ and the overline represents a thermal average, is computed. The
corresponding equation of motion reads
\begin{align}
  z\chi_{ij}(z)=2g\delta_{ij}~\overline{\mu}^z_i&-g\sum_{\ell}\frac{\overline{\mu}^z_i}{\mu_i\mu_\ell} J_{i\ell}~\chi_{\ell j}(z)\nonumber\\&+g\sum_{\ell}\frac{\overline{\mu }^z_\ell}{\mu_i\mu_\ell}J_{i\ell}~\chi_{ij}(z).\label{eqn_susc}
\end{align}
with the electron Land\'e factor $g$ and the energy
$z=E+\text{i}\varepsilon$. 
The disorder is modeled by defining
occupation variables
\begin{align}
  p^{i\alpha}(\vb*{R})=\left\{\begin{array}{lr}
      1&\text{species $\alpha$ on basis site $i$ in unit cell $\vb*{R}$}\\
      0&\text{else}
    \end{array}\right.
\end{align}
and a species resolved Fourier transformation of the susceptibility
\begin{align}
  \chi^{\alpha\beta}_{ij}(\vb*{k},\vb*{k'}):=\sum_{\vb*{R},\vb*{R}'}p^{i\alpha}(\vb*{R})~\text{e}^{-\text{i}\vb*{k}\cdot\vb*{R}}~\chi_{ij}(\vb*{R},\vb*{R'})~p^{j\beta}(\vb*{R'})~\text{e}^{\text{i}\vb*{k'}\cdot\vb*{R'}}.\label{eqn_Ft}
\end{align}
In the following, it is useful to introduce a combined site and
species index denoted by $(i)=i\alpha$, $(j)=j\beta$, etc.. Writing
the susceptibility given in formula \eqref{eqn_susc} as a series and
performing the Fourier transformation (equation \eqref{eqn_Ft}) leads
to expressions with products of Fourier transformed occupation
variables
\begin{align}
  \varrho^{(i)}(\vb*{k})=\sum_{\vb*{R}}p^{(i)}(\vb*{R})~\text{e}^{-\text{i}\vb*{k}\cdot\vb*{R}}.
\end{align}
The averaging process needs to be done very carefully as described in references
\cite{Buczek2016} and \cite{Yonezawa1968} and leads to the appearance
of cumulants of order $n$ given by
\begin{align}
  \mathcal{C}^n_{(\ell_1)(\ell_2)\ldots(\ell_n)}&(\vb*{k}_1,\vb*{k}_2\ldots\vb*{k}_n)=\mathcal{P}^n_{(\ell_1)(\ell_2)\ldots(\ell_n)}(\vb*{c})\nonumber\\&\cdot\varOmega_{\text{BZ}}~\delta(\vb*{k}_1+\vb*{k}_2+\ldots+\vb*{k}_n)
\end{align}
where $\vb*{c}$ is a matrix with the concentrations of each species on
the sublattices and the weight functions
$\mathcal{P}^n_{(\ell_1)(\ell_2)\ldots(\ell_n)}(\vb*{c})$. There is no analytic representation of the latter but the first
two are given by
\begin{align}
  \mathcal{P}^1_{(i)}&=c^{(i)}\nonumber\\
  \mathcal{P}^2_{(i)(j)}&=\delta_{ij}(\delta_{\alpha\beta}~c^{(i)}-c^{(i)}~c^{(j)}).
\end{align}
\begin{figure}
  \includegraphics[width=8cm]{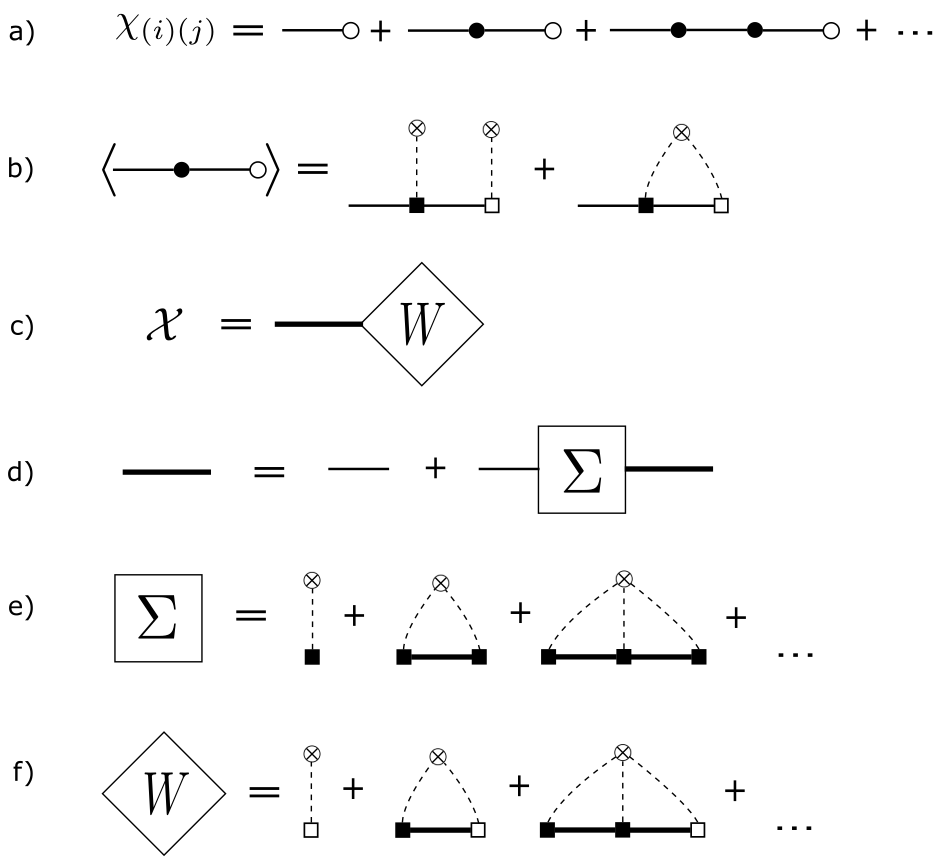}
  \caption{Diagrammatic representation of the main results of the
    CPA-theory. a) Fourier transformation of series \eqref{eqn_susc},
    b) average of the second term in a), c) the averaged
    susceptibility $\mathcal{X}$ written as a product of the effective
    medium propagator $\varXi$ (thick line) and the spin weight $W$, d)
    Dyson equation for the effective medium propagator, e) definition
    of the self-energy $\varSigma$ and f) definition of the spin
    weight $W$}
  \label{fig_feyn}
\end{figure}
A summary of the resulting formulae after the Fourier transformation
and the averaging can be found in figure \ref{fig_feyn} in
diagrammatic form where the following symbols have been used:
\begin{itemize}
\item The $\tau$-matrix
  \begin{align}
    \tau^{(\ell)}_{(i)(j)}(\vb*{k},\vb*{k}')=g\mu_{(j)}^{-1}&\left(J_{(j)(\ell)}(\vb*{k}-\vb*{k'})\frac{\overline{\mu}^z_{(\ell)}}{\mu_{(\ell)}}\delta_{(i)(j)}\right.\nonumber\\&\left.-J_{(\ell)(j)}(\vb*{k}')\frac{\overline{\mu}^z_{(i)}}{\mu_{(i)}}\delta_{(i)(\ell)}\right) 
  \end{align}where
  \begin{align}
    J_{(i)(j)}(\vb*{k})&=\sum_{\vb*{R}}J_{(i)(j)}(\vb*{R})~\text{e}^{-\text{i}\vb*{k}\cdot\vb*{R}}
  \end{align}
  is represented by a filled square.
\item The filled circle represents a $T$- matrix
  \begin{align}
    T_{(i)(j)}(\vb*{k},\vb*{k}')=\sum_{(\ell)}\varrho^{(\ell)}(\vb*{k}-\vb*{k'})~\tau^{(\ell)}_{(i)(j)}(\vb*{k},\vb*{k}').
  \end{align}
\item An empty square stands for a $\sigma$-matrix:
  \begin{align}
    \sigma^{(\ell)}_{(i)(j)}=2g\delta_{(i)(j)}\delta_{(i)(\ell)}\overline{\mu}^z_{(\ell)}
  \end{align}
\item The $S$-matrix is depicted as an empty circle and is given by
  \begin{align}
    S_{(i)(j)}(\vb*{k},\vb*{k'})=\sum_{(\ell)}\varrho^{(\ell)}(\vb*{k}-\vb*{k'})~\sigma^{(\ell)}_{(i)(j)}.
  \end{align}
\item The propagator of uncoupled magnetic moments, represented by a solid line, is given as
  \begin{align}
    \varGamma_{(i)(j)}(z)=z^{-1}\delta_{(i)(j)}.
  \end{align}
\item A cumulant of order $n$ is represented by a crossed
  circle, where the order is given by the number of dashed lines
  ending at it.
\end{itemize}
Furthermore, two rules for the interpretation of the diagrams need to
be followed:
\begin{enumerate}
\item The elements brought together in a diagram undergo a matrix
  multiplication in the $(i)(j)$-space. The corresponding matrix
  indices are written as subscripts in the definitions above.
\item Every internal free propagator is assigned a momentum which is
  integrated over:
  \begin{align}
    \frac{1}{\varOmega_{\text{BZ}}}\int_{\varOmega_{\text{BZ}}}\dd[3]{k_1}
  \end{align}
\end{enumerate}
Every term of the series for the susceptibility in figure
\ref{fig_feyn} a) is averaged independently. The result for the second
term is shown in figure \ref{fig_feyn} b). In the CPA,
crossed terms, which appear in the fourth and higher order terms, are
neglected. This model represents a single-site approximation and
neglects all correlations between two or more sites. As these averaged
diagrams consist of two different vertices (filled and empty squares),
the averaged susceptibility can be written as a product of two
different contributions which we call the effective medium propagator
$\varXi$ and the spin weight $W$ as is shown in figure \ref{fig_feyn} c).
The effective medium propagator is given in terms of a Dyson-equation
shown in figure \ref{fig_feyn} d) with a self-energy defined in figure
\ref{fig_feyn} e). Together with the definition of the spin-weight in
figure \ref{fig_feyn} f), all non-crossed diagrams of the averaged
susceptibility can be constructed.

The calculation of the self-energy is done through the partial
self-energies defined by
\begin{align}
  c^{i\alpha}\hat{\vb*{\varSigma}}^{i\alpha}=\mathcal{P}^1_{i\alpha}\mathds{1}+\mathcal{P}^2_{i\beta,i\alpha}\vb*{M}^{i\beta}+\mathcal{P}^3_{i\gamma,i\beta,i\alpha}\vb*{M}^{i\gamma}\vb*{M}^{i\beta}+\ldots\label{eqn_peps}
\end{align}
where the $M$-matrix is given by
\begin{align}
  \vb*{M}^{(i)}(z,\vb*{k},\vb*{k'})=\vb*{\tau}^{(i)}(\vb*{k},\vb*{k'})~\vb*{\varXi}(z,\vb*{k'}).\label{eqn_Mq}
\end{align}
With that the self-energy is given by 
\begin{align}
  \vb*{\varSigma}(z,\vb*{R},\vb*{R'})=\sum_{(i)}c^{(i)}\sum_{\vb*{R}_1}\hat{\vb*{\varSigma}}^{(i)}(z,\vb*{R},\vb*{R}_1)~\vb*{\tau}^{(i)}(\vb*{R}_1,\vb*{R'})\label{eqn_pawelSlfe}
\end{align}
which can also be seen through its diagrammatic definition. The self-consistency equation inspired by the works of
\cite{Yonezawa1968} and \cite{Matsubara1973} is given by
\begin{align}
  \hat{\vb*{\varSigma}}^{(i)}=\left[\mathds{1}-\left(\vb*{M}^{(i)}-\bar{\vb*{\varSigma}}^i\right)\right]^{-1}\label{eqn_pawelSC}
\end{align} 
where the helping quantity 
\begin{align}
  \bar{\vb*{\varSigma}}^i(\vb*{R},\vb*{R'})=\sum_{\alpha\in I_i}\sum_{\vb*{R}_1}c_{i\alpha}\hat{\vb*{\varSigma}}^{i\alpha}(\vb*{R},\vb*{R}_1)\vb*{M}^{i\alpha}(\vb*{R}_1,\vb*{R'}).\label{eqn_pawelEpsbar}
\end{align}
is used. Equation \eqref{eqn_pawelSC} is used to calculate a new
self-energy from the effective medium propagator with which through
figure \ref{fig_feyn} c) a new effective medium propagator can be
calculated. \\
The temperature dependence is calculated through the
average magnon number
\begin{align}
  \varPhi_{(i)}=\Im{\mathop{\mathlarger{\mathlarger{\int}}}_{-\infty}^{\infty}\dd{z}\frac{D_{(i)}(z)}{\text{e}^{\frac{z}{k_BT}}-1}}\label{eqn_phi}
\end{align}
where
\begin{align}
  D_{(i)}(z)=-\frac{1}{\pi}\mathop{\mathlarger{\mathlarger{\int}}}_{\varOmega_{\text{BZ}}}\dd[3]{k}\frac{\mathcal{X}_{(i)(i)}(z,\vb*{k})}{2gc_{(i)}\overline{\mu}^{z}_{(i)}}.\label{eqn_dos}
\end{align}
Note that the imaginary part of this quantity $D_{(i)}(z)$ is the
magnonic density of states. Following the theory of Callen
\cite{Callen1963} and its implementations in simple disordered and
complex ordered systems \cite{Tang2006,Bouzerar2002,Rusz2005}, the
thermally averaged $z$-component of the magnetic moments are 
\begin{align}
{\scriptstyle
\overline{\mu}^{z}_{(i)}=g\frac{\left(\frac{\mu_{(i)}}{g}-\varPhi_{(i)}\right)\left(1+\varPhi_{(i)}\right)^{\mu_{(i)}+1}+\left(\frac{\mu_{(i)}}{g}+1+\varPhi_{(i)}\right)\varPhi_{(i)}^{\mu_{(i)}+1}}{\left(1+\varPhi_{(i)}\right)^{\mu_{(i)}+1}-\left(\varPhi_{(i)}\right)^{\mu_{(i)}+1}}.\label{eqn_RPA_mu}
}
\end{align}

\section{Implementation}\label{chap_implementation}
\begin{figure}
  \centering
  \includegraphics[width=8cm]{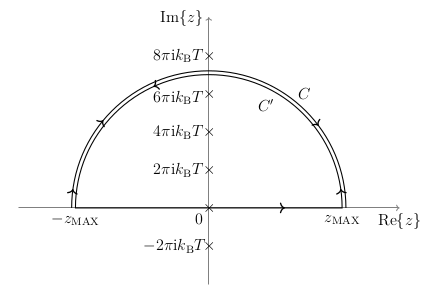}
  \caption{Integration contour used to calculate $\varPhi_{(i)}$. The
    crosses on the imaginary axis mark the poles of
    $\frac{1}{\text{e}^{\beta z}-1}$.}
  \label{fig_contour}
\end{figure}

The integrals in $\vb*{k}$-space (see equation \eqref{eqn_dos}) were
computed using the tetrahedron method \cite{Bloechl1994}. The
energy integral is problematic as $D_{(i)}(z)$ is a rapidly changing
function along the real axis and in addition to that the Bose-factor
$\frac{1}{\text{e}^{\frac{z}{k_{\text{B}}T}}-1}$ has a pole at $z=0$.
Therefore, the energy-integral was implemented using complex contour
integration. The problem was tackled by calculating two complex
integrals, which are shown in figure \ref{fig_contour}. $C$ is a
semi-circle with radius $z_{\text{MAX}}$ and $C'$ is a closed contour
consisting of the same arc as $C$ but in the opposite direction and a
straight line infinitesimally close to the real axis. The closed
contour $C'$ was evaluated using the Residue-theorem as the
Bose-factor has Poles along the imaginary axis at $z_n=2n\pi
\text{i}k_{\text{B}} T$ with $n\in \mathds{Z}$. The values of the
residues are given by
\begin{align}
R(z_n)=k_{\text{B}}TD_{(i)}(z_n)~.
\end{align}
The sum of both contours $C$ and $C'$ gives the integral parallel and
infinitesimally close along the real axis.

This method is based on the fact that the integrand in equation
\eqref{eqn_phi} is analytic almost everywhere in the complex upper
half plane and on the fact that it vanishes for very large positive
and negative energies. The radius of the integration contour
$z_{\text{MAX}}$ was estimated using the Gersgorin disc theorem
\cite{Gersgorin1931}.

Another complication arises from the fact that the Bose-factor has a
singularity at $z=0$. As mentioned above, the method used here gives
the integral parallel to the real axis at an infinitesimal distance
$\varDelta$. Therefore the integral calculated through the complex
contour integral described above is
\begin{align}
\varPhi_{(i)}=\Im{\mathop{\mathlarger{\mathlarger{\int}}}_{-\infty}^{\infty}\frac{D_{(i)}(E+\text{i}\varDelta)}{\text{e}^{\beta E}-1+\text{i}\varDelta\beta\text{e}^{\beta  E}}\dd{E}}
\end{align}
where $\text{e}^{\varDelta x}\approx 1+\varDelta x$ was used. This can be
rewritten using the Shokotski-Plemelj Theorem
\begin{align}
\lim\limits_{\Delta\rightarrow0}\frac{1}{x+\text{i}\Delta}=\frac{\mathcal{P}}{x}-\text{i}\pi\delta(x)\label{eqn_Shokotski}
\end{align}
where $\mathcal{P}$ is the Cauchy principal value. Now, $\varPhi_{(i)}$
is given by the principal value integral but because of the extension
of the integration contour, an additional contribution
\begin{align}
\text{i}\pi k_{\text{B}}TD_{(i)}(0)
\end{align}
is picked up. This contribution is spurious and needs to be subtracted
from the result of the integral.

In the limit $T\rightarrow T_{\text{C}}$, the average magnon number
$\varPhi_{(i)}$ goes to infinity, which allows a series expansion of
equation \eqref{eqn_RPA_mu} in $\frac{1}{\varPhi_{(i)}}$:
\begin{align}
	\overline{\mu}^z_{(i)}=\frac{\mu_{(i)}(\mu_{(i)}+g)}{3g\varPhi_{(i)}}+\order{\left(\frac{1}{\varPhi_{(i)}}\right)^{2}}
\end{align}
Expanding the exponential in formula \eqref{eqn_phi} and inserting it
in the series expansion above leads to
\begin{align}
	\overline{\mu}^z_{(i)}=-\pi\frac{\mu_{(i)}(\mu_{(i)}+g)}{3gk_{\text{B}}T_{\text{C}}}\left[\int\dd{z}\int\dd[3]{k}\frac{\mathcal{X}_{(i)(i)}(z,\vb*{k})}{2gc_{(i)}\overline{\mu}^z_{(i)}z}\right]^{-1}.
\end{align}
An important point is that the latter
equation still holds if all the averaged magnetic moments are
scaled by an arbitrary constant factor. This fact is obvious in
ordered systems as is shown in reference \cite{Rusz2005} and also
holds in substitutionally disordered systems. Using this property, the
calculation of the Curie temperature can be done by treating the
averaged moments as vector and solving the equation
\begin{align}
\overline{\mu}^z_{(i)}=-\pi\frac{\mu_{(i)}(\mu_{(i)}+g)}{3gk_{\text{B}}}\left[\int\dd{z}\int\dd[3]{k}\frac{\mathcal{X}_{(i)(i)}(z,\vb*{k})}{2gc_{(i)}\overline{\mu}^z_{(i)}z}\right]^{-1}.\label{eqn_tc}
\end{align}
iteratively while also normalizing this vector to an arbitrary length
in each step. Note that in equation \eqref{eqn_tc} the factor
$T_{\text{C}}$ is omitted. After convergence is reached, the Curie temperature is given by the length of the vector. 

One of the main advantages of the presented formalism is that the two main
parameters entering the model, magnetic moments $\mu_i^\alpha$ and exchange
constants $J_{ij}$, can be calculated from first-principles. Thus, our
approach in a combination with a density functional theory method
provides a parameter free description of spin waves in substitutional
magnetic alloys and ordered materials at finite temperatures.    

\section{Results}\label{chap_res}
Magnetic moments $\mu_i^\alpha$ and exchange parameters $J_{ij}$ of
iron-cobalt alloys at various concentrations were evaluated using a
first-principles Green-function method within a generalized gradient
approximation of density functional theory~\cite{Perdew1996}. The
method is designed for bulk materials, surfaces, interfaces and
real space
clusters~\cite{Luders2001,Geilhufe2015,Hoffmann2020}. Disorder
effects were taken into account within a coherent-potential
approximation~\cite{Soven1967} as it is implemented within
multiple scattering theory \cite{Gyorffy1972}. The exchange interaction was
estimated using the magnetic force theorem
formulated for substitutional alloys within the CPA
approach~\cite{Turek2006}.

We consider the interaction between 12 shells of neighbors. To
ensure the convergence of calculated properties with the number of neighbor
shells, several calculations were performed for up to 30 shells
showing practically the same results as with 12.

For better comparability, all the results were calculated using a
bcc-structure. Furthermore, the interaction parameters $J_{ij}$ are
held constant (at their value at $T=$0K) while increasing
the temperature.
\subsection{Random disorder}\label{chap_resFeCo}
\subsubsection{Curie temperatures}
\begin{figure}
	\centering
	\includegraphics[width=8cm]{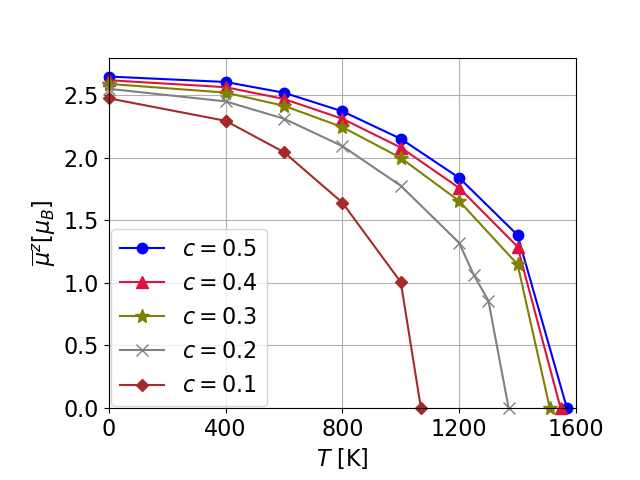}
	\caption{Thermally averaged magnetic moments' $z$-component of iron in \ch{Fe_{1-c}Co_{c}} for different temperatures and cobalt concentrations. Lines are ment as a guide to the eye.}
	\label{fig_TM1}
\end{figure}

\begin{figure}
	\centering
	\includegraphics[width=8cm]{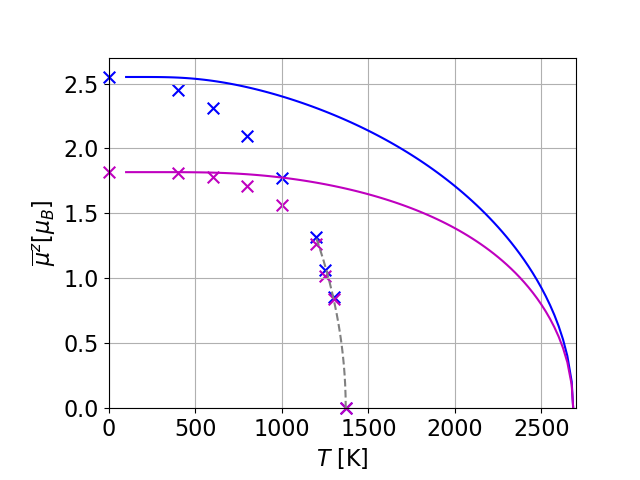}
	\caption{Thermally averaged magnetic moments' z-component in \ch{Fe_{0.8}Co_{0.2}} for different
		temperatures within the RPA (crosses) and the MFA (lines). The
		blue line represents iron while the red line represents
		cobalt. The dashed line represents the expected behavior of the
		Heisenberg model near the Curie temperature.}
	\label{fig_TM2}
\end{figure}

As cobalt has a higher Curie temperature then iron, one would
expect a rise of the critical temperature as the concentration of cobalt
$c$ is increased. Our results shown in figure \ref{fig_TM1} display
this behavior. The points in this figure are the numerical results
which were calculated using the methods described in the previous section.
Near the magnetic phase transition the characteristic behavior of the
averaged magnetic moments is given by
\begin{align}
	\overline{\mu}^z\propto\left(1-\frac{T}{T_\mathrm{C}}\right)^{\beta}.
\end{align} 
The critical exponent $\beta$ has the numeric value of $1/2$ in the
case of the Heisenberg model in the RPA \cite{Kokorina2013}, which is
well known to differ from the experimental value of $\beta\approx1/3$
\cite{Nolting2009}. For the system with $c=0.2$, we used the latter
equation as a fitting function with $\beta=1/2$ for our results close
to $T_{\text{C}}$. It fits very well with our data, cf. figure~\ref{fig_TM2}.
Apart from deploying the RPA, we estimated the Curie temperature
using the mean-field approximation (MFA). The MFA is a purely classical model in
which the thermally averaged magnetic moments are given by
\cite{Nolting2009}
\begin{align}
\overline{\mu}^z_{(i)}=\mu\cdot B_{\mu_{(i)}}\left(\frac{g \mu_{\text{B}}B^m_{(i)}\mu_{(i)}}{k_{\text{B}}T}\right)
\end{align}
with the Bohr magneton $\mu_{\text{B}}$, the Brillouin function $B_\mu(x)$ and the mean field
\begin{align}
B^m_{(i)}=\frac{1}{\mu_{\text{B}}\mu_{(i)}}\sum_{\vb*{R}(j)}J_{(i)(j)}(\vb*{R})c_{(j)}\frac{\overline{\mu}^z_{(j)}}{\mu_{(j)}}.
\end{align}
While the RPA is known to underestimate the Curie temperature
\cite{Rusz2005}, the mean field approximation (MFA) overestimates
it. This is caused by the fact that the MFA neglects the influence of
magnons and therefore only allows spin flips as elementary excitations,
which naturally arises at higher energies than magnons
\cite{Nolting2009}. Thus, the combination of these two methods may be
used to provide bounds for the approximate theoretical
predictions. The MFA equations can be solved iteratively and yield the
results shown in figure 4 for \ch{Fe}$_{0.8}$\ch{Co}$_{0.2}$. They are almost twice
as large as their RPA counterparts, thus providing rather poor account
of the high temperature behavior of the alloy considered.

The results are summarized in table \ref{tab_Curie} together with
experimental results from references
\cite{Normanton1975,Nishizawa1984}.  While RPA performs fairly well, a
clear trend to overestimating the Curie temperature can be seen.
Partially, the behavior can be attributed to the fact that in our
calculations we restrict the system to a bcc-lattice, while the
real iron cobalt system will undergo a structural phase transition at
elevated temperatures \cite{Nishizawa1984} which is expected to influence the
Curie temperature.

\begin{table}
  \centering
  \begin{tabular}{|c|c|c|c|c|}
    \hline
    $c$&$T_{\text{C}}^{\text{RPA}}~[\unit{K}]$&$T_{\text{C}}^{\text{MFA}}~[\unit{K}]$&$T_{\text{C}}~[\unit{K}]$ in \cite{Nishizawa1984}&$T_{\text{C}}~[\unit{K}]$ in \cite{Normanton1975}\\\hline
    0.1&1069&2199&1164&1144\\\hline
    0.2&1369&2684&1225&1211\\\hline
    0.3&1510&2844&1260&1243\\\hline
    0.4&1547&2837&1268&1250\\\hline
    0.5&1568&2803&1265&1243\\\hline
  \end{tabular}
  \caption{Comparison of the Curie temperatures calculated in this work with experimental results in \cite{Normanton1975,Nishizawa1984}}
 \label{tab_Curie}
 \end{table}


\subsubsection{Magnonic spectrum}
\begin{figure}
  \centering
  \includegraphics[width=8cm]{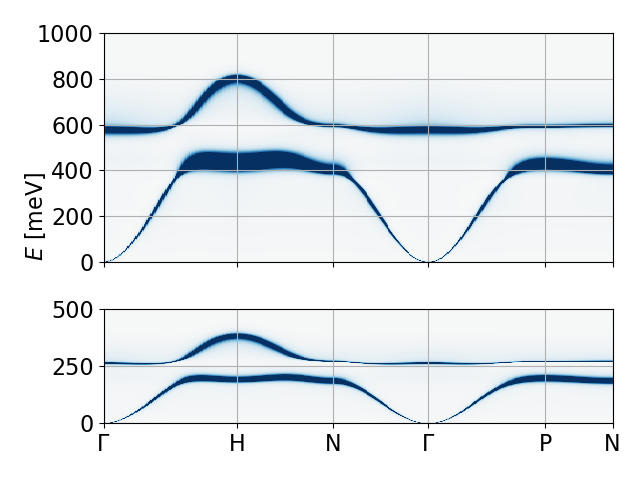}
  \caption{Magnonic spectrum (trace of the imaginary part of the averaged transverse susceptibility) of \ch{Fe_{0.8}Co_{0.2}} at $T=$0\ K (top) and $T\approx0.9\ T_{\text{C}}$ (bottom).}
  \label{fig_FeCo}
\end{figure}

We extract the magnonic spectrum from the imaginary part of the
retarded averaged susceptibility by calculating its trace $\sum_{(i)}\vb*{\mathcal{X}}_{(i)(i)}(z,\vb*{k})$. The most prominent
feature is its bandgap appearing due to strongly different interaction
strengths and magnetic moments between different constituents.  Our
results suggest that this bandgap is stable up to high temperatures as
can be seen in the result for \ch{Fe_{0.8}Co_{0.2}} presented in
figure \ref{fig_FeCo}.  In the upper (lower) plot the spectrum for the
case of $T=$0\ K ($T\approx0.9 T_{\text{C}}$) is shown.
Interestingly, the main features of the band structure are preserved
as the temperature increases.  The scaling (softening) of the magnonic
spectrum propositional to the thermally averaged magnetic moment is a
feature of the RPA.  In this approximation, the magnon energies vanish
above the Curie temperature. In a more sophisticated treatment, the
spectrum above the critical temperature should feature paramagnetic
like excitations emerging as a manifestation of the short-range
magnetic order \cite{Essenberger2012}.

Let us note that the peaks feature finite width appearing due to the
presence of disorder in the system. The damping is relatively small
and increases somehwat only at elevated energies, in particular close
to the edges of the bandgap. Within the RPA, the width of the peaks is independent of temperature as can be seen from equation \ref{eqn_susc}.

\subsubsection{Width of the bandgap, spin stiffness and lifetimes}
\begin{figure}
    \includegraphics[width=8cm]{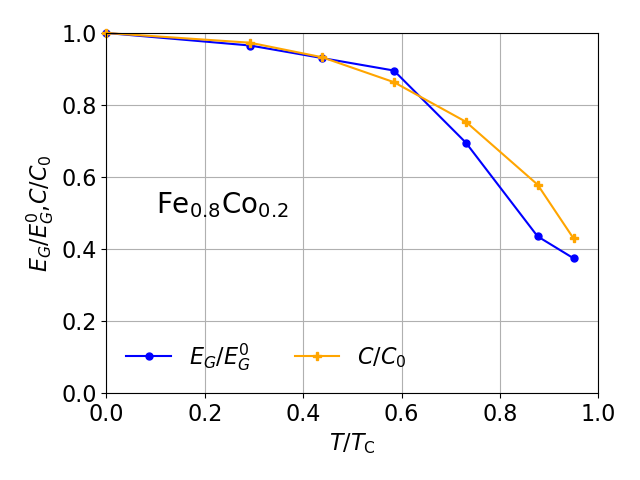}
    \includegraphics[width=8cm]{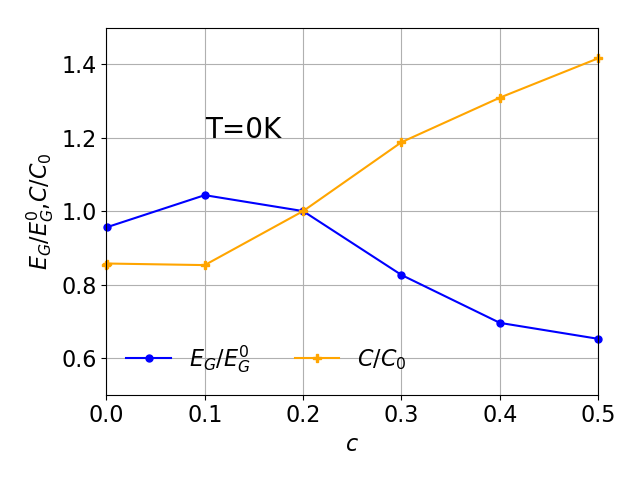}
  \caption{Relative width of the bandgap $\frac{E_G}{E_G^0}$ and
    relative spin stiffness $\frac{C}{C_0}$ for different
    Cobalt-concentrations at \\$T=$0\ K (bottom) and for different
    temperatures at $c=20\%$ (top). All quantities are
    normalized to their values at $c=20\%$ and
    $T=$0\ K ($C_0\approx477\text{meV}$\AA$^2$,
    $E_G^0=115\text{meV}$).}
	\label{fig_gapC}
\end{figure}

We investigate the spin wave stiffness constant $C$ describing the
quadratic magnon dispersion of the acoustic mode in the long
wave-length limit
\begin{align}
E\ = Ck^{2}
\end{align}
as well as the size of the bandgap.  Both decrease roughly
proportionally to the average magnetization (see figure
\ref{fig_gapC}) as the temperature is increased. The reference values at at $c=20\%$ and
$T=$0\ K
\begin{align}
	C_0\approx 477\text{meV\AA}^2\qquad E_G^0=115\text{meV}
\end{align}
are in reasonable agreement with values from other studies of iron and cobalt \cite{Buczek2011,Hueller1986}.

Furthermore, we determine the full width at half maximum (FWHM) of the
magnon peaks for several wave-vectors. The FWHM is computed using a
Lorentzian fit function for the imaginary part of the susceptibility
as function of the energy:
\begin{align}
\Im{\chi}(E)\approx h\frac{\frac{1}{2}\text{FWHM}}{(E-E_0)^2+\left(\frac{1}{2}\text{FWHM}\right)^2}
\end{align}
with the location of the maximum $E_0$ of the peak with scaling factor
$h$.  The FWHM is interpreted as the inverse
magnon lifetime.  In order to facilitate an quantitative comparison,
we normalize the width to the energy of the magnon for a particular
wave-vector.  This feature can be interpreted as the inverse of the
quality factor, giving the amount of energy leaking from the mode per
cycle of the precession.

\begin{figure}
    \includegraphics[width=8cm]{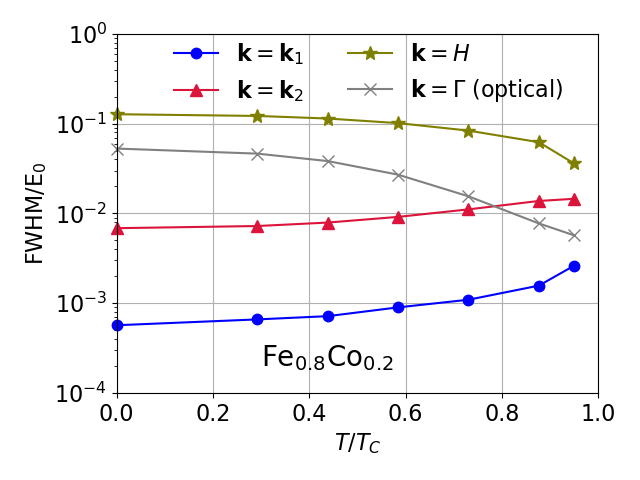}
    \includegraphics[width=8cm]{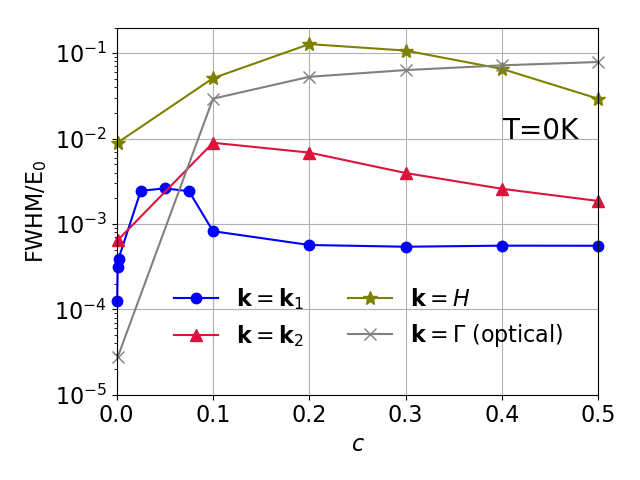}
  \caption{FWHM at
    $\vb*{k}_1=(0.1,0,0)\frac{1}{a_{\text{B}}}$,
    $\vb*{k}_2=(0.57,0,0)\frac{1}{a_{\text{B}}}$ (midway
    between $\Gamma$ and H), H and the optical mode at
    $\Gamma$ for different Cobalt-concentrations at
    $T=$0\ K (bottom) and for different temperatures at
    $c=20\%$ (top). The FWHM is normalized by the magnon energy $E_0$ at the corresponding temperature and concentration.}
	\label{fig_fwhmC}
\end{figure}

We recall that in our formalism the finite widths of the magnon
resonances arise only due to the action of the disorder.
Nevertheless, at constant Co concentration $c$, the FWHM varies with
the temperature as well.  In a simple picture, this somewhat
unexpected observation can be interpreted as follows:  The
scattering rate of magnons of particular energy on the crystal
imperfections (or alternatively the FWHM for weak coupling) is
proportional to the concentration of dopants and the density of final
magnon states with this energy, as the scattering potential is static.
Even though the density of states decreases with
temperature, it does not necessarily retain its shape.  Thus, for
different modes with different wave-vectors the density of available
finite states will vary as the temperature is raised.  As evident from figure \ref{fig_fwhmC}, this effect depends on the magnon state.
With rising temperature, the normalized widths increase for low
energy acoustic magnons, but decrease for magnons at the top of the
acoustic branch and in the optical branch.

However, we note again that our prediction concerning the evolution
of the width with the temperature, due to the use of the RPA, does not
include the main mechanism, i.e. the coupling of the magnons to the
thermal bath.  In the RPA, without disorder, the magnons would
feature an infinite lifetime.  In general, it is expected, that the
thermally induced width should increase with the temperature
\cite{Knoll1990}.\\
The disorder induced broadening of the magnon peaks in the alloys studied here is found to be in general smaller than $50$meV. Other studies of similar ordered systems which include Landau damping, generally estimate much higher damping. {\c{S}}a{\c{s}}{\i}o{\u{g}}lu \emph{et. al.} \cite{Sasiouglu2013} study tetragonal \ch{FeCo} compounds predicting acoustic magnon modes with widths between $50$meV and $100$meV and optic modes with widths between $60$meV and $200$meV at the edges of the Brillouin zone based on MBT. Buczek et. al. \cite{Buczek2011a} predict widths of more than $100$meV for high energy modes in bulk fcc \ch{Co}, and more than $60$meV in the case of bulf bcc iron based on TDDFT calculations. They also report spin wave disappearance in bcc iron close to the H point with widths as high as $550$meV in that region. Consequently, we come to the conclusion that the disorder induced damping is rather small compared to Landau damping in the considered systems.

\begin{figure}
  \centering
  \includegraphics[width=8cm]{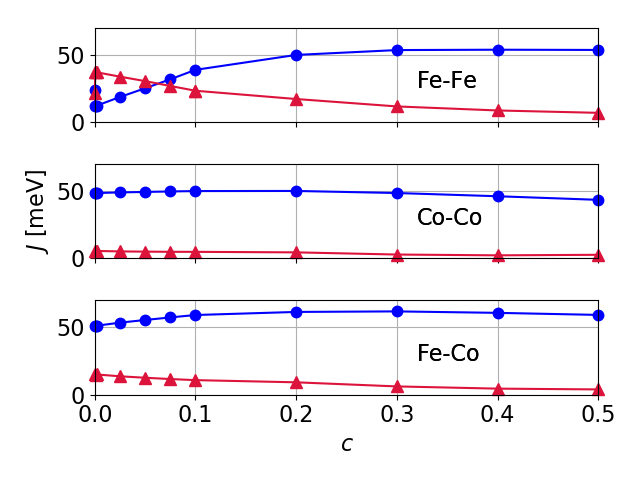}
  \caption{Exchange interaction in iron-cobalt alloys nearest neighbors (blue circles), next nearest neighbors (red triangles). The interaction with atoms in outer shells is comparably weak.}
  \label{fig_J}
\end{figure}

\begin{figure*}
  \centering
    \includegraphics[width=8cm]{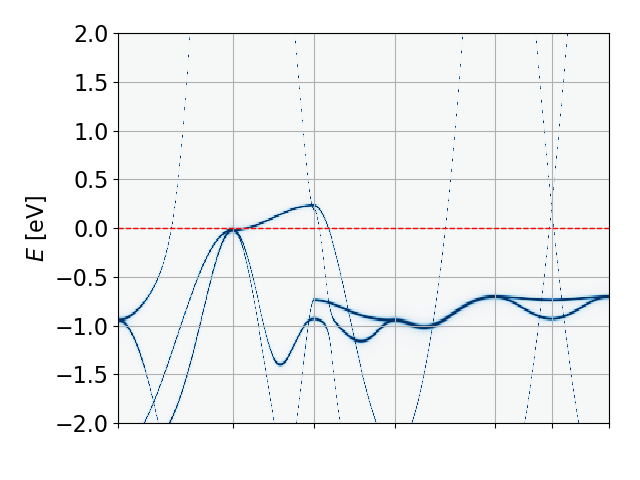}
    \includegraphics[width=8cm]{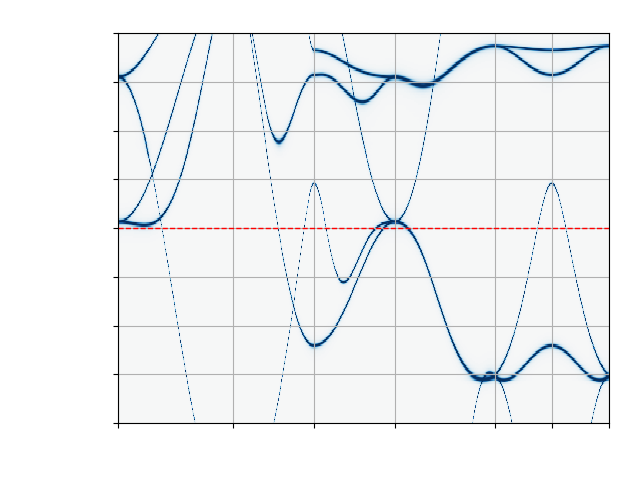}
    \includegraphics[width=8cm]{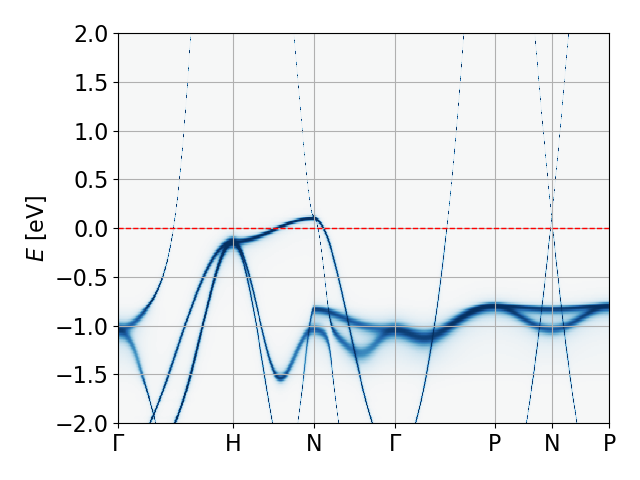}
    \includegraphics[width=8cm]{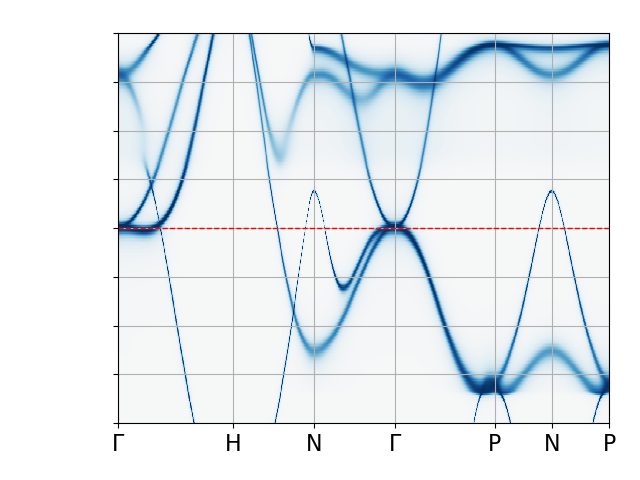}
  \caption{Bloch spectral functions for
    \ch{Fe}$_{0.995}$\ch{Co}$_{0.005}$ (upper panels) and
    \ch{Fe}$_{0.9}$\ch{Co}$_{0.1}$ (lower panels) for majority (left) and
    minority spin channels, respectively. The red dotted line represents the Fermi energy.}
  \label{fig_BSF}
\end{figure*}

\begin{figure}
	\centering
	\includegraphics[width=8cm]{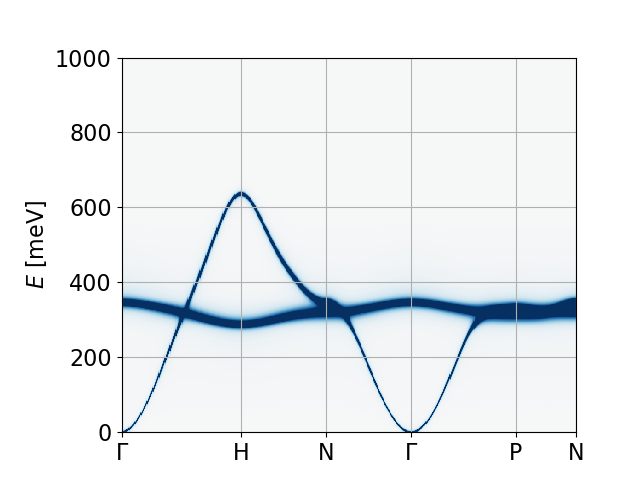}
	\caption{Magnonic spectrum of \ch{Fe_{0.5}Co_{0.5}} at $T=$0\ K and $\mu_{\text{Fe}}=\mu_{\text{Co}}$.}
	\label{fig_FeCo2}
\end{figure}

The evolution of the magnonic spectrum with the disorder shows several
interesting features.  For small Co concentrations, the bandgap
increases slightly and above $c \approx 0.1$ starts to decrease with
$c$, cf. figure \ref{fig_gapC}.  As mentioned before, in simple terms,
the gap arises because of the large difference in the exchange
integrals (magnetic interactions) and magnetic moments between
different constituents. Figure \ref{fig_J} shows that this
difference is pronounced most strongly for low concentrations.  The
strong increase of the nearest neighbor \ch{Fe}-\ch{Fe} interaction as
the \ch{Co} concentration increases causes the bandgap to get
narrower, as this exchange integral becomes similar in magnitude to
the \ch{Co}-\ch{Co} interaction. The enhancement of \ch{Fe}-\ch{Fe}
exchange interaction with increase of \ch{Co} concentration can be
explained by a strong hybridization between $3d$ states of \ch{Fe} and
\ch{Co} atoms. In addition, the presence of Co leads to enhancement of
the density of states at the Fermi level, increasing the Stoner factor
and the exchange interaction. Increase of the Co concentration fills
up the bands mainly in the minority spin channel. Fig.~\ref{fig_BSF}
shows the calculated electronic band structure (Bloch spectral function) for $c=0.5\%$ and
$c=10\%$ for both majority and minority spin channels,
respectively. The most important changes for different Co concentrations occur along the $\Gamma$-H
line for the majority bands and in the vicinity of the $\Gamma$ point for the minority bands. At low Co
concentrations a band along the $\Gamma$-H is in the Fermi level's
vicinity but is not occupied. It is filled up at higher Co
concentrations ($c>5\%$) and leads to a significant increase of the
magnetic interaction in the systems.  At high cobalt
concentrations, it is mainly the difference of the magnetic moments
which prevents the closing of the bandgap. To verify this statement we
show the spectrum of \ch{Fe_{0.5}Co_{0.5}} with equal magnetic moments
for both constituents $\mu_{\text{Fe}}=\mu_{\text{Co}}$. As can be
seen in figure \ref{fig_FeCo2}, the bandgap closes in this case.

\begin{figure}
	\centering
	\includegraphics[width=8cm]{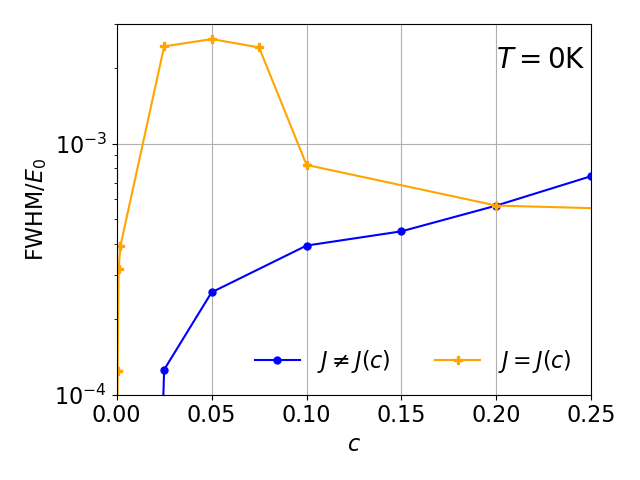}
	\caption{FWHM/$E_0$ for different concentrations if the
          interaction parameters are held constant (blue circles) and
          if they change with the cobalt concentration (orange
          crosses). The FWHM is normalized by the magnon energy $E_0$ at $T=0$K and the corresponding concentration.}
	\label{fig_fwhm}
\end{figure}

Finally, we note that the FWHM shows maxima at certain concentrations,
which are caused by the change of the exchange parameters. In figure
\ref{fig_fwhm}, we show the FWHM at
$\vb*{k}_1=(0.1,0,0)\frac{1}{a_{\text{B}}}$ and
$T=$0\ K for different concentrations compared to the FWHM for
the case of fixed interactions.

\subsection{Short range order}\label{chap_resFeCo_sro}
\begin{figure}
    \includegraphics[width=8cm]{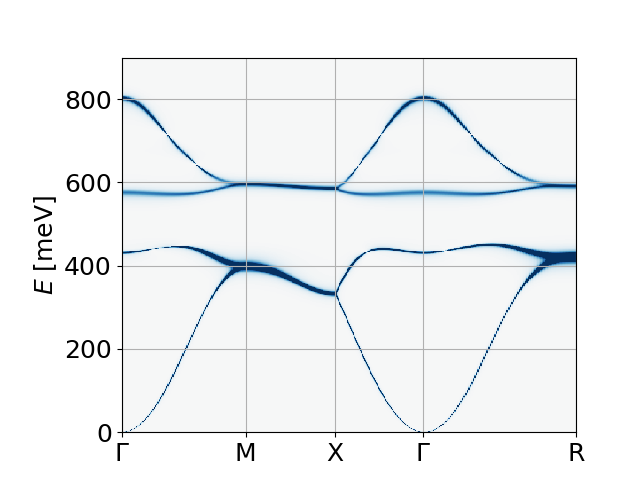}
    \includegraphics[width=8cm]{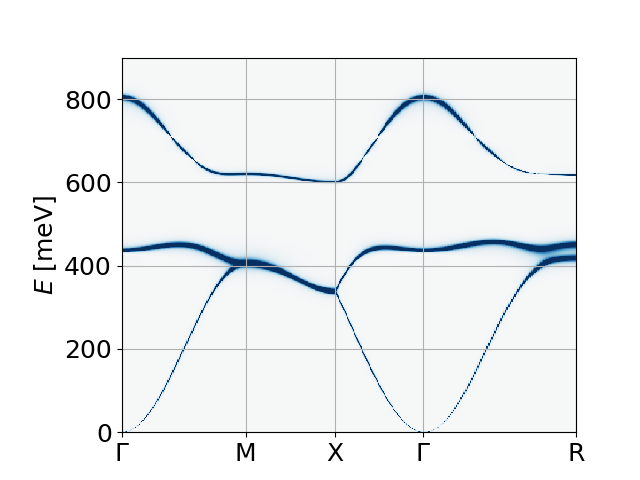}
  \caption{Magnonic spectrum of \ch{Fe_{0.8}Co_{0.2}} with random disorder (top), and in a configuration in which
    all cobalt atoms are isolated from each other according to
    the occupation probabilities given in table \ref{tab_sro} at
    $T=$0\ K
    (bottom).
  }
  \label{fig_sro}
\end{figure}

Our theory is formulated in the framework of the single-site CPA,
which by definition is not able to account for the appearance of short
range order or any other correlations between the occupation of
different sites. However, through our generalization of the theory to
lattices with multiple atoms per unit cell, we are able to include
short range order through different occupation probabilities within
the unit cell. In this section, we discuss the influence of short range
order using a very simple model. Instead of performing the
calculations for the primitive unit cell, we choose for the case of an
alloy exhibiting short range order the usage of the cubic unit cell
with 2 atoms and the occupation probabilities listed in table
\ref{tab_sro}.
\begin{table}
	\centering
	\begin{tabular}{|c|c|c|}
		\hline
		element&site 1&site 2\\\hline
		\ch{Fe}&1&0.6\\\hline
		\ch{Co}&0&0.4\\\hline
	\end{tabular}
	\caption{Occupation probabilities for the case of short range order.}
	\label{tab_sro}
\end{table}
This configuration corresponds to an alloy in which two cobalt atoms
never sit next to each other. 
The results for the case of random disorder and short range order are compared in figure \ref{fig_sro}.
As there are now two basis sites occupied with
two elements according to table \ref{tab_sro}, the spectrum now
consists of three bands. The main result of this test is the
verification that the bandgap remains present in the case of an alloy
exhibiting short range order.

The magnonic properties discussed above in the alloy with SRO
compute to
\begin{align}
	E_G&\approx 115\text{meV}\nonumber\\
	\frac{C}{C_0}&=1.03\\
	\frac{\text{FWHM}}{\text{FWHM}_0}&=1.92.\nonumber
\end{align}
It can be seen that the width of the bandgap and the spin stiffness
hardly change at all, but the FWHM nearly doubles its value.
Obviously, this is far from a complete study of the influence of SRO,
but it suggests that the inclusion of SRO will only have a minor
impact on the width of the bandgap.

\section{Summary}\label{chap_sum}
We presented a first-principle approach to calculate
critical magnetic phenomena and spin waves at finite temperatures for
complex disordered materials. The method is based on a mapping of a
Green function, obtained within the  multiple scattering theory, on
the Heisenberg model using a coherent potential approximation.  The
temperature effects were taken into account within an RPA for
the magnonic Green function. 

Our approach is illustrated on disordered iron-cobalt alloys which
exhibit many of the properties demanded from magnonic crystals: They
exhibit a bandgap whose width shows an interesting behavior in the
concentration and temperature range studied in this work. The
influence of short range order on the bandgap turns out to be of minor
importance in our calculations. However, the latter result should only
be seen as an intermediate step obtained for one specific type of
SRO and needs further investigation.

The temperature dependence of the bandwidth and the spin stiffness mirrors the decreasing magnetization as the temperature is
increased. Thus the treatment of temperature is far from complete. 
Moreover, the inclusion of Landau damping in the description of disordered systems is a further necessary improvement of the theory which we currently develop. 
\bibliographystyle{apsrev4-2}
\bibliography{./Quellen}

\begin{thebibliography}{62}%
\makeatletter
\providecommand \@ifxundefined [1]{%
 \@ifx{#1\undefined}
}%
\providecommand \@ifnum [1]{%
 \ifnum #1\expandafter \@firstoftwo
 \else \expandafter \@secondoftwo
 \fi
}%
\providecommand \@ifx [1]{%
 \ifx #1\expandafter \@firstoftwo
 \else \expandafter \@secondoftwo
 \fi
}%
\providecommand \natexlab [1]{#1}%
\providecommand \enquote  [1]{``#1''}%
\providecommand \bibnamefont  [1]{#1}%
\providecommand \bibfnamefont [1]{#1}%
\providecommand \citenamefont [1]{#1}%
\providecommand \href@noop [0]{\@secondoftwo}%
\providecommand \href [0]{\begingroup \@sanitize@url \@href}%
\providecommand \@href[1]{\@@startlink{#1}\@@href}%
\providecommand \@@href[1]{\endgroup#1\@@endlink}%
\providecommand \@sanitize@url [0]{\catcode `\\12\catcode `\$12\catcode
  `\&12\catcode `\#12\catcode `\^12\catcode `\_12\catcode `\%12\relax}%
\providecommand \@@startlink[1]{}%
\providecommand \@@endlink[0]{}%
\providecommand \url  [0]{\begingroup\@sanitize@url \@url }%
\providecommand \@url [1]{\endgroup\@href {#1}{\urlprefix }}%
\providecommand \urlprefix  [0]{URL }%
\providecommand \Eprint [0]{\href }%
\providecommand \doibase [0]{https://doi.org/}%
\providecommand \selectlanguage [0]{\@gobble}%
\providecommand \bibinfo  [0]{\@secondoftwo}%
\providecommand \bibfield  [0]{\@secondoftwo}%
\providecommand \translation [1]{[#1]}%
\providecommand \BibitemOpen [0]{}%
\providecommand \bibitemStop [0]{}%
\providecommand \bibitemNoStop [0]{.\EOS\space}%
\providecommand \EOS [0]{\spacefactor3000\relax}%
\providecommand \BibitemShut  [1]{\csname bibitem#1\endcsname}%
\let\auto@bib@innerbib\@empty
\bibitem [{\citenamefont {Chumak}\ \emph {et~al.}(2015)\citenamefont {Chumak},
  \citenamefont {Vasyuchka}, \citenamefont {Serga},\ and\ \citenamefont
  {Hillebrands}}]{Chumak2015}%
  \BibitemOpen
  \bibfield  {author} {\bibinfo {author} {\bibfnamefont {A.~V.}\ \bibnamefont
  {Chumak}}, \bibinfo {author} {\bibfnamefont {V.~I.}\ \bibnamefont
  {Vasyuchka}}, \bibinfo {author} {\bibfnamefont {A.~A.}\ \bibnamefont
  {Serga}},\ and\ \bibinfo {author} {\bibfnamefont {B.}~\bibnamefont
  {Hillebrands}},\ }\href {https://doi.org/10.1038/nphys3347} {\bibfield
  {journal} {\bibinfo  {journal} {Nature Physics}\ }\textbf {\bibinfo {volume}
  {11}},\ \bibinfo {pages} {453} (\bibinfo {year} {2015})}\BibitemShut
  {NoStop}%
\bibitem [{\citenamefont {Khitun}\ and\ \citenamefont
  {Wang}(2011)}]{Khitun2011}%
  \BibitemOpen
  \bibfield  {author} {\bibinfo {author} {\bibfnamefont {A.}~\bibnamefont
  {Khitun}}\ and\ \bibinfo {author} {\bibfnamefont {K.~L.}\ \bibnamefont
  {Wang}},\ }\href {https://doi.org/10.1063/1.3609062} {\bibfield  {journal}
  {\bibinfo  {journal} {Journal of Applied Physics}\ }\textbf {\bibinfo
  {volume} {110}},\ \bibinfo {pages} {034306} (\bibinfo {year}
  {2011})}\BibitemShut {NoStop}%
\bibitem [{\citenamefont {Al-Wahsh}\ \emph {et~al.}(2011)\citenamefont
  {Al-Wahsh}, \citenamefont {Akjouj}, \citenamefont {Djafari-Rouhani},\ and\
  \citenamefont {Dobrzynski}}]{AlWahsh2011}%
  \BibitemOpen
  \bibfield  {author} {\bibinfo {author} {\bibfnamefont {H.}~\bibnamefont
  {Al-Wahsh}}, \bibinfo {author} {\bibfnamefont {A.}~\bibnamefont {Akjouj}},
  \bibinfo {author} {\bibfnamefont {B.}~\bibnamefont {Djafari-Rouhani}},\ and\
  \bibinfo {author} {\bibfnamefont {L.}~\bibnamefont {Dobrzynski}},\ }\href
  {https://doi.org/10.1016/j.surfrep.2010.10.002} {\bibfield  {journal}
  {\bibinfo  {journal} {Surface Science Reports}\ }\textbf {\bibinfo {volume}
  {66}},\ \bibinfo {pages} {29} (\bibinfo {year} {2011})}\BibitemShut {NoStop}%
\bibitem [{\citenamefont {Bloch}(1930)}]{Bloch1930}%
  \BibitemOpen
  \bibfield  {author} {\bibinfo {author} {\bibfnamefont {F.}~\bibnamefont
  {Bloch}},\ }\href {https://doi.org/10.1007/bf01339661} {\bibfield  {journal}
  {\bibinfo  {journal} {Zeitschrift f\"ur Physik}\ }\textbf {\bibinfo {volume}
  {61}},\ \bibinfo {pages} {206} (\bibinfo {year} {1930})}\BibitemShut
  {NoStop}%
\bibitem [{\citenamefont {Chumak}\ \emph {et~al.}(2017)\citenamefont {Chumak},
  \citenamefont {Serga},\ and\ \citenamefont {Hillebrands}}]{Chumak2017}%
  \BibitemOpen
  \bibfield  {author} {\bibinfo {author} {\bibfnamefont {A.~V.}\ \bibnamefont
  {Chumak}}, \bibinfo {author} {\bibfnamefont {A.~A.}\ \bibnamefont {Serga}},\
  and\ \bibinfo {author} {\bibfnamefont {B.}~\bibnamefont {Hillebrands}},\
  }\href {https://doi.org/10.1088/1361-6463/aa6a65} {\bibfield  {journal}
  {\bibinfo  {journal} {Journal of Physics D: Applied Physics}\ }\textbf
  {\bibinfo {volume} {50}},\ \bibinfo {pages} {244001} (\bibinfo {year}
  {2017})}\BibitemShut {NoStop}%
\bibitem [{\citenamefont {Nikitov}\ \emph {et~al.}(2001)\citenamefont
  {Nikitov}, \citenamefont {Tailhades},\ and\ \citenamefont
  {Tsai}}]{Nikitov2001}%
  \BibitemOpen
  \bibfield  {author} {\bibinfo {author} {\bibfnamefont {S.}~\bibnamefont
  {Nikitov}}, \bibinfo {author} {\bibfnamefont {P.}~\bibnamefont {Tailhades}},\
  and\ \bibinfo {author} {\bibfnamefont {C.}~\bibnamefont {Tsai}},\ }\href
  {https://doi.org/10.1016/s0304-8853(01)00470-x} {\bibfield  {journal}
  {\bibinfo  {journal} {Journal of Magnetism and Magnetic Materials}\ }\textbf
  {\bibinfo {volume} {236}},\ \bibinfo {pages} {320} (\bibinfo {year}
  {2001})}\BibitemShut {NoStop}%
\bibitem [{\citenamefont {Krawczyk}\ and\ \citenamefont
  {Grundler}(2014)}]{Krawczyk2014}%
  \BibitemOpen
  \bibfield  {author} {\bibinfo {author} {\bibfnamefont {M.}~\bibnamefont
  {Krawczyk}}\ and\ \bibinfo {author} {\bibfnamefont {D.}~\bibnamefont
  {Grundler}},\ }\href {https://doi.org/10.1088/0953-8984/26/12/123202}
  {\bibfield  {journal} {\bibinfo  {journal} {Journal of Physics: Condensed
  Matter}\ }\textbf {\bibinfo {volume} {26}},\ \bibinfo {pages} {123202}
  (\bibinfo {year} {2014})}\BibitemShut {NoStop}%
\bibitem [{\citenamefont {Lenk}\ \emph {et~al.}(2011)\citenamefont {Lenk},
  \citenamefont {Ulrichs}, \citenamefont {Garbs},\ and\ \citenamefont
  {Münzenberg}}]{Lenk2011}%
  \BibitemOpen
  \bibfield  {author} {\bibinfo {author} {\bibfnamefont {B.}~\bibnamefont
  {Lenk}}, \bibinfo {author} {\bibfnamefont {H.}~\bibnamefont {Ulrichs}},
  \bibinfo {author} {\bibfnamefont {F.}~\bibnamefont {Garbs}},\ and\ \bibinfo
  {author} {\bibfnamefont {M.}~\bibnamefont {Münzenberg}},\ }\href
  {https://doi.org/10.1016/j.physrep.2011.06.003} {\bibfield  {journal}
  {\bibinfo  {journal} {Physics Reports}\ }\textbf {\bibinfo {volume} {507}},\
  \bibinfo {pages} {107} (\bibinfo {year} {2011})}\BibitemShut {NoStop}%
\bibitem [{\citenamefont {Sadovnikov}\ \emph {et~al.}(2016)\citenamefont
  {Sadovnikov}, \citenamefont {Beginin}, \citenamefont {Odincov}, \citenamefont
  {Sheshukova}, \citenamefont {Sharaevskii}, \citenamefont {Stognij},\ and\
  \citenamefont {Nikitov}}]{Sadovnikov2016}%
  \BibitemOpen
  \bibfield  {author} {\bibinfo {author} {\bibfnamefont {A.~V.}\ \bibnamefont
  {Sadovnikov}}, \bibinfo {author} {\bibfnamefont {E.~N.}\ \bibnamefont
  {Beginin}}, \bibinfo {author} {\bibfnamefont {S.~A.}\ \bibnamefont
  {Odincov}}, \bibinfo {author} {\bibfnamefont {S.~E.}\ \bibnamefont
  {Sheshukova}}, \bibinfo {author} {\bibfnamefont {Y.~P.}\ \bibnamefont
  {Sharaevskii}}, \bibinfo {author} {\bibfnamefont {A.~I.}\ \bibnamefont
  {Stognij}},\ and\ \bibinfo {author} {\bibfnamefont {S.~A.}\ \bibnamefont
  {Nikitov}},\ }\href {https://doi.org/10.1063/1.4948381} {\bibfield  {journal}
  {\bibinfo  {journal} {Applied Physics Letters}\ }\textbf {\bibinfo {volume}
  {108}},\ \bibinfo {pages} {172411} (\bibinfo {year} {2016})}\BibitemShut
  {NoStop}%
\bibitem [{\citenamefont {Sadovnikov}\ \emph {et~al.}(2018)\citenamefont
  {Sadovnikov}, \citenamefont {Gubanov}, \citenamefont {Sheshukova},
  \citenamefont {Sharaevskii},\ and\ \citenamefont {Nikitov}}]{Sadovnikov2018}%
  \BibitemOpen
  \bibfield  {author} {\bibinfo {author} {\bibfnamefont {A.~V.}\ \bibnamefont
  {Sadovnikov}}, \bibinfo {author} {\bibfnamefont {V.~A.}\ \bibnamefont
  {Gubanov}}, \bibinfo {author} {\bibfnamefont {S.~E.}\ \bibnamefont
  {Sheshukova}}, \bibinfo {author} {\bibfnamefont {Y.~P.}\ \bibnamefont
  {Sharaevskii}},\ and\ \bibinfo {author} {\bibfnamefont {S.~A.}\ \bibnamefont
  {Nikitov}},\ }\bibfield  {journal} {\bibinfo  {journal} {Physical Review
  Applied}\ }\textbf {\bibinfo {volume} {9}},\ \href
  {https://doi.org/10.1103/physrevapplied.9.051002}
  {10.1103/physrevapplied.9.051002} (\bibinfo {year} {2018})\BibitemShut
  {NoStop}%
\bibitem [{\citenamefont {Sheshukova}\ \emph {et~al.}(2013)\citenamefont
  {Sheshukova}, \citenamefont {Morozova}, \citenamefont {Beginin},
  \citenamefont {Sharaevskii},\ and\ \citenamefont {Nikitov}}]{Sheshukova2013}%
  \BibitemOpen
  \bibfield  {author} {\bibinfo {author} {\bibfnamefont {S.~E.}\ \bibnamefont
  {Sheshukova}}, \bibinfo {author} {\bibfnamefont {M.~A.}\ \bibnamefont
  {Morozova}}, \bibinfo {author} {\bibfnamefont {E.~N.}\ \bibnamefont
  {Beginin}}, \bibinfo {author} {\bibfnamefont {Y.~P.}\ \bibnamefont
  {Sharaevskii}},\ and\ \bibinfo {author} {\bibfnamefont {S.~A.}\ \bibnamefont
  {Nikitov}},\ }\href {https://doi.org/10.3103/s1541308x13040134} {\bibfield
  {journal} {\bibinfo  {journal} {Physics of Wave Phenomena}\ }\textbf
  {\bibinfo {volume} {21}},\ \bibinfo {pages} {304} (\bibinfo {year}
  {2013})}\BibitemShut {NoStop}%
\bibitem [{\citenamefont {Zakeri}(2018)}]{Zakeri2018}%
  \BibitemOpen
  \bibfield  {author} {\bibinfo {author} {\bibfnamefont {K.}~\bibnamefont
  {Zakeri}},\ }\href {https://doi.org/10.1016/j.physc.2018.02.035} {\bibfield
  {journal} {\bibinfo  {journal} {Physica C: Superconductivity and its
  Applications}\ }\textbf {\bibinfo {volume} {549}},\ \bibinfo {pages} {164}
  (\bibinfo {year} {2018})}\BibitemShut {NoStop}%
\bibitem [{\citenamefont {Buczek}\ \emph {et~al.}(2010)\citenamefont {Buczek},
  \citenamefont {Ernst},\ and\ \citenamefont {Sandratskii}}]{Buczek2010d}%
  \BibitemOpen
  \bibfield  {author} {\bibinfo {author} {\bibfnamefont {P.}~\bibnamefont
  {Buczek}}, \bibinfo {author} {\bibfnamefont {A.}~\bibnamefont {Ernst}},\ and\
  \bibinfo {author} {\bibfnamefont {L.~M.}\ \bibnamefont {Sandratskii}},\
  }\bibfield  {journal} {\bibinfo  {journal} {Physical Review Letters}\
  }\textbf {\bibinfo {volume} {105}},\ \href
  {https://doi.org/10.1103/physrevlett.105.097205}
  {10.1103/physrevlett.105.097205} (\bibinfo {year} {2010})\BibitemShut
  {NoStop}%
\bibitem [{\citenamefont {Buczek}\ \emph {et~al.}(2009)\citenamefont {Buczek},
  \citenamefont {Ernst}, \citenamefont {Bruno},\ and\ \citenamefont
  {Sandratskii}}]{Buczek2009}%
  \BibitemOpen
  \bibfield  {author} {\bibinfo {author} {\bibfnamefont {P.}~\bibnamefont
  {Buczek}}, \bibinfo {author} {\bibfnamefont {A.}~\bibnamefont {Ernst}},
  \bibinfo {author} {\bibfnamefont {P.}~\bibnamefont {Bruno}},\ and\ \bibinfo
  {author} {\bibfnamefont {L.~M.}\ \bibnamefont {Sandratskii}},\ }\bibfield
  {journal} {\bibinfo  {journal} {Physical Review Letters}\ }\textbf {\bibinfo
  {volume} {102}},\ \href {https://doi.org/10.1103/physrevlett.102.247206}
  {10.1103/physrevlett.102.247206} (\bibinfo {year} {2009})\BibitemShut
  {NoStop}%
\bibitem [{\citenamefont {Normanton}\ \emph {et~al.}(1975)\citenamefont
  {Normanton}, \citenamefont {Bloomfield}, \citenamefont {Sale},\ and\
  \citenamefont {Argent}}]{Normanton1975}%
  \BibitemOpen
  \bibfield  {author} {\bibinfo {author} {\bibfnamefont {A.~S.}\ \bibnamefont
  {Normanton}}, \bibinfo {author} {\bibfnamefont {P.~E.}\ \bibnamefont
  {Bloomfield}}, \bibinfo {author} {\bibfnamefont {F.~R.}\ \bibnamefont
  {Sale}},\ and\ \bibinfo {author} {\bibfnamefont {B.~B.}\ \bibnamefont
  {Argent}},\ }\href {https://doi.org/10.1179/030634575790444658} {\bibfield
  {journal} {\bibinfo  {journal} {Metal Science}\ }\textbf {\bibinfo {volume}
  {9}},\ \bibinfo {pages} {510} (\bibinfo {year} {1975})}\BibitemShut {NoStop}%
\bibitem [{\citenamefont {Nishizawa}\ and\ \citenamefont
  {Ishida}(1984)}]{Nishizawa1984}%
  \BibitemOpen
  \bibfield  {author} {\bibinfo {author} {\bibfnamefont {T.}~\bibnamefont
  {Nishizawa}}\ and\ \bibinfo {author} {\bibfnamefont {K.}~\bibnamefont
  {Ishida}},\ }\href {https://doi.org/10.1007/bf02868548} {\bibfield  {journal}
  {\bibinfo  {journal} {Bulletin of Alloy Phase Diagrams}\ }\textbf {\bibinfo
  {volume} {5}},\ \bibinfo {pages} {250} (\bibinfo {year} {1984})}\BibitemShut
  {NoStop}%
\bibitem [{\citenamefont {Buczek}\ \emph
  {et~al.}(2011{\natexlab{a}})\citenamefont {Buczek}, \citenamefont {Ernst},\
  and\ \citenamefont {Sandratskii}}]{Buczek2011}%
  \BibitemOpen
  \bibfield  {author} {\bibinfo {author} {\bibfnamefont {P.}~\bibnamefont
  {Buczek}}, \bibinfo {author} {\bibfnamefont {A.}~\bibnamefont {Ernst}},\ and\
  \bibinfo {author} {\bibfnamefont {L.~M.}\ \bibnamefont {Sandratskii}},\
  }\bibfield  {journal} {\bibinfo  {journal} {Physical Review Letters}\
  }\textbf {\bibinfo {volume} {106}},\ \href
  {https://doi.org/10.1103/physrevlett.106.157204}
  {10.1103/physrevlett.106.157204} (\bibinfo {year}
  {2011}{\natexlab{a}})\BibitemShut {NoStop}%
\bibitem [{\citenamefont {Buczek}\ \emph
  {et~al.}(2011{\natexlab{b}})\citenamefont {Buczek}, \citenamefont {Ernst},\
  and\ \citenamefont {Sandratskii}}]{Buczek2011a}%
  \BibitemOpen
  \bibfield  {author} {\bibinfo {author} {\bibfnamefont {P.}~\bibnamefont
  {Buczek}}, \bibinfo {author} {\bibfnamefont {A.}~\bibnamefont {Ernst}},\ and\
  \bibinfo {author} {\bibfnamefont {L.~M.}\ \bibnamefont {Sandratskii}},\
  }\bibfield  {journal} {\bibinfo  {journal} {Physical Review B}\ }\textbf
  {\bibinfo {volume} {84}},\ \href {https://doi.org/10.1103/physrevb.84.174418}
  {10.1103/physrevb.84.174418} (\bibinfo {year}
  {2011}{\natexlab{b}})\BibitemShut {NoStop}%
\bibitem [{\citenamefont {Zakeri}\ \emph {et~al.}(2012)\citenamefont {Zakeri},
  \citenamefont {Zhang}, \citenamefont {Chuang},\ and\ \citenamefont
  {Kirschner}}]{Zhang2012}%
  \BibitemOpen
  \bibfield  {author} {\bibinfo {author} {\bibfnamefont {K.}~\bibnamefont
  {Zakeri}}, \bibinfo {author} {\bibfnamefont {Y.}~\bibnamefont {Zhang}},
  \bibinfo {author} {\bibfnamefont {T.-H.}\ \bibnamefont {Chuang}},\ and\
  \bibinfo {author} {\bibfnamefont {J.}~\bibnamefont {Kirschner}},\ }\bibfield
  {journal} {\bibinfo  {journal} {Physical Review Letters}\ }\textbf {\bibinfo
  {volume} {108}},\ \href {https://doi.org/10.1103/physrevlett.108.197205}
  {10.1103/physrevlett.108.197205} (\bibinfo {year} {2012})\BibitemShut
  {NoStop}%
\bibitem [{\citenamefont {Qin}\ \emph {et~al.}(2015)\citenamefont {Qin},
  \citenamefont {Zakeri}, \citenamefont {Ernst}, \citenamefont {Sandratskii},
  \citenamefont {Buczek}, \citenamefont {Marmodoro}, \citenamefont {Chuang},
  \citenamefont {Zhang},\ and\ \citenamefont {Kirschner}}]{Qin2015}%
  \BibitemOpen
  \bibfield  {author} {\bibinfo {author} {\bibfnamefont {H.~J.}\ \bibnamefont
  {Qin}}, \bibinfo {author} {\bibfnamefont {K.}~\bibnamefont {Zakeri}},
  \bibinfo {author} {\bibfnamefont {A.}~\bibnamefont {Ernst}}, \bibinfo
  {author} {\bibfnamefont {L.~M.}\ \bibnamefont {Sandratskii}}, \bibinfo
  {author} {\bibfnamefont {P.}~\bibnamefont {Buczek}}, \bibinfo {author}
  {\bibfnamefont {A.}~\bibnamefont {Marmodoro}}, \bibinfo {author}
  {\bibfnamefont {T.~H.}\ \bibnamefont {Chuang}}, \bibinfo {author}
  {\bibfnamefont {Y.}~\bibnamefont {Zhang}},\ and\ \bibinfo {author}
  {\bibfnamefont {J.}~\bibnamefont {Kirschner}},\ }\bibfield  {journal}
  {\bibinfo  {journal} {Nature Communications}\ }\textbf {\bibinfo {volume}
  {6}},\ \href {https://doi.org/10.1038/ncomms7126} {10.1038/ncomms7126}
  (\bibinfo {year} {2015})\BibitemShut {NoStop}%
\bibitem [{\citenamefont {Buczek}\ \emph {et~al.}(2016)\citenamefont {Buczek},
  \citenamefont {Sandratskii}, \citenamefont {Buczek}, \citenamefont {Thomas},
  \citenamefont {Vignale},\ and\ \citenamefont {Ernst}}]{Buczek2016}%
  \BibitemOpen
  \bibfield  {author} {\bibinfo {author} {\bibfnamefont {P.}~\bibnamefont
  {Buczek}}, \bibinfo {author} {\bibfnamefont {L.~M.}\ \bibnamefont
  {Sandratskii}}, \bibinfo {author} {\bibfnamefont {N.}~\bibnamefont {Buczek}},
  \bibinfo {author} {\bibfnamefont {S.}~\bibnamefont {Thomas}}, \bibinfo
  {author} {\bibfnamefont {G.}~\bibnamefont {Vignale}},\ and\ \bibinfo {author}
  {\bibfnamefont {A.}~\bibnamefont {Ernst}},\ }\bibfield  {journal} {\bibinfo
  {journal} {Physical Review B}\ }\textbf {\bibinfo {volume} {94}},\ \href
  {https://doi.org/10.1103/physrevb.94.054407} {10.1103/physrevb.94.054407}
  (\bibinfo {year} {2016})\BibitemShut {NoStop}%
\bibitem [{\citenamefont {Buczek}\ \emph {et~al.}(2018)\citenamefont {Buczek},
  \citenamefont {Thomas}, \citenamefont {Marmodoro}, \citenamefont {Buczek},
  \citenamefont {Zubizarreta}, \citenamefont {Hoffmann}, \citenamefont
  {Balashov}, \citenamefont {Wulfhekel}, \citenamefont {Zakeri},\ and\
  \citenamefont {Ernst}}]{Buczek2018}%
  \BibitemOpen
  \bibfield  {author} {\bibinfo {author} {\bibfnamefont {P.}~\bibnamefont
  {Buczek}}, \bibinfo {author} {\bibfnamefont {S.}~\bibnamefont {Thomas}},
  \bibinfo {author} {\bibfnamefont {A.}~\bibnamefont {Marmodoro}}, \bibinfo
  {author} {\bibfnamefont {N.}~\bibnamefont {Buczek}}, \bibinfo {author}
  {\bibfnamefont {X.}~\bibnamefont {Zubizarreta}}, \bibinfo {author}
  {\bibfnamefont {M.}~\bibnamefont {Hoffmann}}, \bibinfo {author}
  {\bibfnamefont {T.}~\bibnamefont {Balashov}}, \bibinfo {author}
  {\bibfnamefont {W.}~\bibnamefont {Wulfhekel}}, \bibinfo {author}
  {\bibfnamefont {K.}~\bibnamefont {Zakeri}},\ and\ \bibinfo {author}
  {\bibfnamefont {A.}~\bibnamefont {Ernst}},\ }\href
  {https://doi.org/10.1088/1361-648x/aadefb} {\bibfield  {journal} {\bibinfo
  {journal} {Journal of Physics: Condensed Matter}\ }\textbf {\bibinfo {volume}
  {30}},\ \bibinfo {pages} {423001} (\bibinfo {year} {2018})}\BibitemShut
  {NoStop}%
\bibitem [{\citenamefont {Zakeri}(2014)}]{Zakeri2014}%
  \BibitemOpen
  \bibfield  {author} {\bibinfo {author} {\bibfnamefont {K.}~\bibnamefont
  {Zakeri}},\ }\href {https://doi.org/10.1016/j.physrep.2014.08.001} {\bibfield
   {journal} {\bibinfo  {journal} {Physics Reports}\ }\textbf {\bibinfo
  {volume} {545}},\ \bibinfo {pages} {47} (\bibinfo {year} {2014})}\BibitemShut
  {NoStop}%
\bibitem [{\citenamefont {Halilov}\ \emph {et~al.}(1998)\citenamefont
  {Halilov}, \citenamefont {Eschrig}, \citenamefont {Perlov},\ and\
  \citenamefont {Oppeneer}}]{Halilov_1998}%
  \BibitemOpen
  \bibfield  {author} {\bibinfo {author} {\bibfnamefont {S.~V.}\ \bibnamefont
  {Halilov}}, \bibinfo {author} {\bibfnamefont {H.}~\bibnamefont {Eschrig}},
  \bibinfo {author} {\bibfnamefont {A.~Y.}\ \bibnamefont {Perlov}},\ and\
  \bibinfo {author} {\bibfnamefont {P.~M.}\ \bibnamefont {Oppeneer}},\ }\href
  {https://doi.org/10.1103/physrevb.58.293} {\bibfield  {journal} {\bibinfo
  {journal} {Physical Review B}\ }\textbf {\bibinfo {volume} {58}},\ \bibinfo
  {pages} {293} (\bibinfo {year} {1998})}\BibitemShut {NoStop}%
\bibitem [{\citenamefont {Etz}\ \emph {et~al.}(2015)\citenamefont {Etz},
  \citenamefont {Bergqvist}, \citenamefont {Bergman}, \citenamefont {Taroni},\
  and\ \citenamefont {Eriksson}}]{Etz2015}%
  \BibitemOpen
  \bibfield  {author} {\bibinfo {author} {\bibfnamefont {C.}~\bibnamefont
  {Etz}}, \bibinfo {author} {\bibfnamefont {L.}~\bibnamefont {Bergqvist}},
  \bibinfo {author} {\bibfnamefont {A.}~\bibnamefont {Bergman}}, \bibinfo
  {author} {\bibfnamefont {A.}~\bibnamefont {Taroni}},\ and\ \bibinfo {author}
  {\bibfnamefont {O.}~\bibnamefont {Eriksson}},\ }\href
  {https://doi.org/10.1088/0953-8984/27/24/243202} {\bibfield  {journal}
  {\bibinfo  {journal} {Journal of Physics: Condensed Matter}\ }\textbf
  {\bibinfo {volume} {27}},\ \bibinfo {pages} {243202} (\bibinfo {year}
  {2015})}\BibitemShut {NoStop}%
\bibitem [{\citenamefont {Pajda}\ \emph {et~al.}(2001)\citenamefont {Pajda},
  \citenamefont {Kudrnovsk{\'{y}}}, \citenamefont {Turek}, \citenamefont
  {Drchal},\ and\ \citenamefont {Bruno}}]{Pajda2001}%
  \BibitemOpen
  \bibfield  {author} {\bibinfo {author} {\bibfnamefont {M.}~\bibnamefont
  {Pajda}}, \bibinfo {author} {\bibfnamefont {J.}~\bibnamefont
  {Kudrnovsk{\'{y}}}}, \bibinfo {author} {\bibfnamefont {I.}~\bibnamefont
  {Turek}}, \bibinfo {author} {\bibfnamefont {V.}~\bibnamefont {Drchal}},\ and\
  \bibinfo {author} {\bibfnamefont {P.}~\bibnamefont {Bruno}},\ }\bibfield
  {journal} {\bibinfo  {journal} {Physical Review B}\ }\textbf {\bibinfo
  {volume} {64}},\ \href {https://doi.org/10.1103/physrevb.64.174402}
  {10.1103/physrevb.64.174402} (\bibinfo {year} {2001})\BibitemShut {NoStop}%
\bibitem [{\citenamefont {Callen}(1963)}]{Callen1963}%
  \BibitemOpen
  \bibfield  {author} {\bibinfo {author} {\bibfnamefont {H.~B.}\ \bibnamefont
  {Callen}},\ }\href {https://doi.org/10.1103/physrev.130.890} {\bibfield
  {journal} {\bibinfo  {journal} {Physical Review}\ }\textbf {\bibinfo {volume}
  {130}},\ \bibinfo {pages} {890} (\bibinfo {year} {1963})}\BibitemShut
  {NoStop}%
\bibitem [{\citenamefont {{\c{S}}a{\c{s}}{\i}o{\u{g}}lu}\ \emph
  {et~al.}(2013{\natexlab{a}})\citenamefont {{\c{S}}a{\c{s}}{\i}o{\u{g}}lu},
  \citenamefont {Friedrich},\ and\ \citenamefont {Blügel}}]{Bluegel2013}%
  \BibitemOpen
  \bibfield  {author} {\bibinfo {author} {\bibfnamefont {E.}~\bibnamefont
  {{\c{S}}a{\c{s}}{\i}o{\u{g}}lu}}, \bibinfo {author} {\bibfnamefont
  {C.}~\bibnamefont {Friedrich}},\ and\ \bibinfo {author} {\bibfnamefont
  {S.}~\bibnamefont {Blügel}},\ }\bibfield  {journal} {\bibinfo  {journal}
  {Physical Review B}\ }\textbf {\bibinfo {volume} {87}},\ \href
  {https://doi.org/10.1103/physrevb.87.020410} {10.1103/physrevb.87.020410}
  (\bibinfo {year} {2013}{\natexlab{a}})\BibitemShut {NoStop}%
\bibitem [{\citenamefont {Okumura}\ \emph {et~al.}(2019)\citenamefont
  {Okumura}, \citenamefont {Sato},\ and\ \citenamefont {Kotani}}]{Okumura2019}%
  \BibitemOpen
  \bibfield  {author} {\bibinfo {author} {\bibfnamefont {H.}~\bibnamefont
  {Okumura}}, \bibinfo {author} {\bibfnamefont {K.}~\bibnamefont {Sato}},\ and\
  \bibinfo {author} {\bibfnamefont {T.}~\bibnamefont {Kotani}},\ }\href
  {https://doi.org/10.1103/physrevb.100.054419} {\bibfield  {journal} {\bibinfo
   {journal} {Physical Review B}\ }\textbf {\bibinfo {volume} {100}},\ \bibinfo
  {pages} {054419} (\bibinfo {year} {2019})}\BibitemShut {NoStop}%
\bibitem [{\citenamefont {{\c{S}}a{\c{s}}{\i}o{\u{g}}lu}\ \emph
  {et~al.}(2013{\natexlab{b}})\citenamefont {{\c{S}}a{\c{s}}{\i}o{\u{g}}lu},
  \citenamefont {Friedrich},\ and\ \citenamefont {Blügel}}]{Sasiouglu2013}%
  \BibitemOpen
  \bibfield  {author} {\bibinfo {author} {\bibfnamefont {E.}~\bibnamefont
  {{\c{S}}a{\c{s}}{\i}o{\u{g}}lu}}, \bibinfo {author} {\bibfnamefont
  {C.}~\bibnamefont {Friedrich}},\ and\ \bibinfo {author} {\bibfnamefont
  {S.}~\bibnamefont {Blügel}},\ }\bibfield  {journal} {\bibinfo  {journal}
  {Physical Review B}\ }\textbf {\bibinfo {volume} {87}},\ \href
  {https://doi.org/10.1103/physrevb.87.020410} {10.1103/physrevb.87.020410}
  (\bibinfo {year} {2013}{\natexlab{b}})\BibitemShut {NoStop}%
\bibitem [{\citenamefont {Zhang}\ \emph {et~al.}(2010)\citenamefont {Zhang},
  \citenamefont {Buczek}, \citenamefont {Sandratskii}, \citenamefont {Tang},
  \citenamefont {Prokop}, \citenamefont {Tudosa}, \citenamefont {Peixoto},
  \citenamefont {Zakeri},\ and\ \citenamefont {Kirschner}}]{Zhang2010}%
  \BibitemOpen
  \bibfield  {author} {\bibinfo {author} {\bibfnamefont {Y.}~\bibnamefont
  {Zhang}}, \bibinfo {author} {\bibfnamefont {P.}~\bibnamefont {Buczek}},
  \bibinfo {author} {\bibfnamefont {L.}~\bibnamefont {Sandratskii}}, \bibinfo
  {author} {\bibfnamefont {W.~X.}\ \bibnamefont {Tang}}, \bibinfo {author}
  {\bibfnamefont {J.}~\bibnamefont {Prokop}}, \bibinfo {author} {\bibfnamefont
  {I.}~\bibnamefont {Tudosa}}, \bibinfo {author} {\bibfnamefont {T.~R.~F.}\
  \bibnamefont {Peixoto}}, \bibinfo {author} {\bibfnamefont {K.}~\bibnamefont
  {Zakeri}},\ and\ \bibinfo {author} {\bibfnamefont {J.}~\bibnamefont
  {Kirschner}},\ }\bibfield  {journal} {\bibinfo  {journal} {Physical Review
  B}\ }\textbf {\bibinfo {volume} {81}},\ \href
  {https://doi.org/10.1103/physrevb.81.094438} {10.1103/physrevb.81.094438}
  (\bibinfo {year} {2010})\BibitemShut {NoStop}%
\bibitem [{\citenamefont {Etzkorn}\ \emph {et~al.}(2005)\citenamefont
  {Etzkorn}, \citenamefont {Kumar}, \citenamefont {Tang}, \citenamefont
  {Zhang},\ and\ \citenamefont {Kirschner}}]{Etzkorn2005}%
  \BibitemOpen
  \bibfield  {author} {\bibinfo {author} {\bibfnamefont {M.}~\bibnamefont
  {Etzkorn}}, \bibinfo {author} {\bibfnamefont {P.~S.~A.}\ \bibnamefont
  {Kumar}}, \bibinfo {author} {\bibfnamefont {W.}~\bibnamefont {Tang}},
  \bibinfo {author} {\bibfnamefont {Y.}~\bibnamefont {Zhang}},\ and\ \bibinfo
  {author} {\bibfnamefont {J.}~\bibnamefont {Kirschner}},\ }\bibfield
  {journal} {\bibinfo  {journal} {Physical Review B}\ }\textbf {\bibinfo
  {volume} {72}},\ \href {https://doi.org/10.1103/physrevb.72.184420}
  {10.1103/physrevb.72.184420} (\bibinfo {year} {2005})\BibitemShut {NoStop}%
\bibitem [{\citenamefont {Tang}\ \emph {et~al.}(2007)\citenamefont {Tang},
  \citenamefont {Zhang}, \citenamefont {Tudosa}, \citenamefont {Prokop},
  \citenamefont {Etzkorn},\ and\ \citenamefont {Kirschner}}]{Tang2007}%
  \BibitemOpen
  \bibfield  {author} {\bibinfo {author} {\bibfnamefont {W.~X.}\ \bibnamefont
  {Tang}}, \bibinfo {author} {\bibfnamefont {Y.}~\bibnamefont {Zhang}},
  \bibinfo {author} {\bibfnamefont {I.}~\bibnamefont {Tudosa}}, \bibinfo
  {author} {\bibfnamefont {J.}~\bibnamefont {Prokop}}, \bibinfo {author}
  {\bibfnamefont {M.}~\bibnamefont {Etzkorn}},\ and\ \bibinfo {author}
  {\bibfnamefont {J.}~\bibnamefont {Kirschner}},\ }\bibfield  {journal}
  {\bibinfo  {journal} {Physical Review Letters}\ }\textbf {\bibinfo {volume}
  {99}},\ \href {https://doi.org/10.1103/physrevlett.99.087202}
  {10.1103/physrevlett.99.087202} (\bibinfo {year} {2007})\BibitemShut
  {NoStop}%
\bibitem [{\citenamefont {Vollmer}\ \emph {et~al.}(2003)\citenamefont
  {Vollmer}, \citenamefont {Etzkorn}, \citenamefont {Kumar}, \citenamefont
  {Ibach},\ and\ \citenamefont {Kirschner}}]{Vollmer2003}%
  \BibitemOpen
  \bibfield  {author} {\bibinfo {author} {\bibfnamefont {R.}~\bibnamefont
  {Vollmer}}, \bibinfo {author} {\bibfnamefont {M.}~\bibnamefont {Etzkorn}},
  \bibinfo {author} {\bibfnamefont {P.~S.~A.}\ \bibnamefont {Kumar}}, \bibinfo
  {author} {\bibfnamefont {H.}~\bibnamefont {Ibach}},\ and\ \bibinfo {author}
  {\bibfnamefont {J.}~\bibnamefont {Kirschner}},\ }\bibfield  {journal}
  {\bibinfo  {journal} {Physical Review Letters}\ }\textbf {\bibinfo {volume}
  {91}},\ \href {https://doi.org/10.1103/physrevlett.91.147201}
  {10.1103/physrevlett.91.147201} (\bibinfo {year} {2003})\BibitemShut
  {NoStop}%
\bibitem [{\citenamefont {Meng}\ \emph {et~al.}(2014)\citenamefont {Meng},
  \citenamefont {Zakeri}, \citenamefont {Ernst}, \citenamefont {Chuang},
  \citenamefont {Qin}, \citenamefont {Chen},\ and\ \citenamefont
  {Kirschner}}]{Meng2014}%
  \BibitemOpen
  \bibfield  {author} {\bibinfo {author} {\bibfnamefont {Y.}~\bibnamefont
  {Meng}}, \bibinfo {author} {\bibfnamefont {K.}~\bibnamefont {Zakeri}},
  \bibinfo {author} {\bibfnamefont {A.}~\bibnamefont {Ernst}}, \bibinfo
  {author} {\bibfnamefont {T.-H.}\ \bibnamefont {Chuang}}, \bibinfo {author}
  {\bibfnamefont {H.~J.}\ \bibnamefont {Qin}}, \bibinfo {author} {\bibfnamefont
  {Y.-J.}\ \bibnamefont {Chen}},\ and\ \bibinfo {author} {\bibfnamefont
  {J.}~\bibnamefont {Kirschner}},\ }\href
  {https://doi.org/10.1103/physrevb.90.174437} {\bibfield  {journal} {\bibinfo
  {journal} {Physical Review B}\ }\textbf {\bibinfo {volume} {90}},\ \bibinfo
  {pages} {174437} (\bibinfo {year} {2014})}\BibitemShut {NoStop}%
\bibitem [{\citenamefont {Chuang}\ \emph {et~al.}(2014)\citenamefont {Chuang},
  \citenamefont {Zakeri}, \citenamefont {Ernst}, \citenamefont {Zhang},
  \citenamefont {Qin}, \citenamefont {Meng}, \citenamefont {Chen},\ and\
  \citenamefont {Kirschner}}]{Chuang2014}%
  \BibitemOpen
  \bibfield  {author} {\bibinfo {author} {\bibfnamefont {T.-H.}\ \bibnamefont
  {Chuang}}, \bibinfo {author} {\bibfnamefont {K.}~\bibnamefont {Zakeri}},
  \bibinfo {author} {\bibfnamefont {A.}~\bibnamefont {Ernst}}, \bibinfo
  {author} {\bibfnamefont {Y.}~\bibnamefont {Zhang}}, \bibinfo {author}
  {\bibfnamefont {H.~J.}\ \bibnamefont {Qin}}, \bibinfo {author} {\bibfnamefont
  {Y.}~\bibnamefont {Meng}}, \bibinfo {author} {\bibfnamefont {Y.-J.}\
  \bibnamefont {Chen}},\ and\ \bibinfo {author} {\bibfnamefont
  {J.}~\bibnamefont {Kirschner}},\ }\href
  {https://doi.org/10.1103/physrevb.89.174404} {\bibfield  {journal} {\bibinfo
  {journal} {Physical Review B}\ }\textbf {\bibinfo {volume} {89}},\ \bibinfo
  {pages} {174404} (\bibinfo {year} {2014})}\BibitemShut {NoStop}%
\bibitem [{\citenamefont {Zakeri}\ \emph {et~al.}(2013)\citenamefont {Zakeri},
  \citenamefont {Chuang}, \citenamefont {Ernst}, \citenamefont {Sandratskii},
  \citenamefont {Buczek}, \citenamefont {Qin}, \citenamefont {Zhang},\ and\
  \citenamefont {Kirschner}}]{Zakeri2013}%
  \BibitemOpen
  \bibfield  {author} {\bibinfo {author} {\bibfnamefont {K.}~\bibnamefont
  {Zakeri}}, \bibinfo {author} {\bibfnamefont {T.-H.}\ \bibnamefont {Chuang}},
  \bibinfo {author} {\bibfnamefont {A.}~\bibnamefont {Ernst}}, \bibinfo
  {author} {\bibfnamefont {L.~M.}\ \bibnamefont {Sandratskii}}, \bibinfo
  {author} {\bibfnamefont {P.}~\bibnamefont {Buczek}}, \bibinfo {author}
  {\bibfnamefont {H.~J.}\ \bibnamefont {Qin}}, \bibinfo {author} {\bibfnamefont
  {Y.}~\bibnamefont {Zhang}},\ and\ \bibinfo {author} {\bibfnamefont
  {J.}~\bibnamefont {Kirschner}},\ }\href
  {https://doi.org/10.1038/nnano.2013.188} {\bibfield  {journal} {\bibinfo
  {journal} {Nature Nanotechnology}\ }\textbf {\bibinfo {volume} {8}},\
  \bibinfo {pages} {853} (\bibinfo {year} {2013})}\BibitemShut {NoStop}%
\bibitem [{\citenamefont {Qin}\ \emph {et~al.}(2019)\citenamefont {Qin},
  \citenamefont {Tsurkan}, \citenamefont {Ernst},\ and\ \citenamefont
  {Zakeri}}]{Qin2019}%
  \BibitemOpen
  \bibfield  {author} {\bibinfo {author} {\bibfnamefont {H.}~\bibnamefont
  {Qin}}, \bibinfo {author} {\bibfnamefont {S.}~\bibnamefont {Tsurkan}},
  \bibinfo {author} {\bibfnamefont {A.}~\bibnamefont {Ernst}},\ and\ \bibinfo
  {author} {\bibfnamefont {K.}~\bibnamefont {Zakeri}},\ }\href
  {https://doi.org/10.1103/physrevlett.123.257202} {\bibfield  {journal}
  {\bibinfo  {journal} {Physical Review Letters}\ }\textbf {\bibinfo {volume}
  {123}},\ \bibinfo {pages} {257202} (\bibinfo {year} {2019})}\BibitemShut
  {NoStop}%
\bibitem [{\citenamefont {Zakeri}\ \emph {et~al.}(2021)\citenamefont {Zakeri},
  \citenamefont {Qin},\ and\ \citenamefont {Ernst}}]{Zakeri2021}%
  \BibitemOpen
  \bibfield  {author} {\bibinfo {author} {\bibfnamefont {K.}~\bibnamefont
  {Zakeri}}, \bibinfo {author} {\bibfnamefont {H.}~\bibnamefont {Qin}},\ and\
  \bibinfo {author} {\bibfnamefont {A.}~\bibnamefont {Ernst}},\ }\bibfield
  {journal} {\bibinfo  {journal} {Communications Physics}\ }\textbf {\bibinfo
  {volume} {4}},\ \href {https://doi.org/10.1038/s42005-021-00521-7}
  {10.1038/s42005-021-00521-7} (\bibinfo {year} {2021})\BibitemShut {NoStop}%
\bibitem [{\citenamefont {Qin}\ \emph {et~al.}(2013)\citenamefont {Qin},
  \citenamefont {Zakeri}, \citenamefont {Ernst}, \citenamefont {Chuang},
  \citenamefont {Chen}, \citenamefont {Meng},\ and\ \citenamefont
  {Kirschner}}]{Qin2013}%
  \BibitemOpen
  \bibfield  {author} {\bibinfo {author} {\bibfnamefont {H.~J.}\ \bibnamefont
  {Qin}}, \bibinfo {author} {\bibfnamefont {K.}~\bibnamefont {Zakeri}},
  \bibinfo {author} {\bibfnamefont {A.}~\bibnamefont {Ernst}}, \bibinfo
  {author} {\bibfnamefont {T.-H.}\ \bibnamefont {Chuang}}, \bibinfo {author}
  {\bibfnamefont {Y.-J.}\ \bibnamefont {Chen}}, \bibinfo {author}
  {\bibfnamefont {Y.}~\bibnamefont {Meng}},\ and\ \bibinfo {author}
  {\bibfnamefont {J.}~\bibnamefont {Kirschner}},\ }\bibfield  {journal}
  {\bibinfo  {journal} {Physical Review B}\ }\textbf {\bibinfo {volume} {88}},\
  \href {https://doi.org/10.1103/physrevb.88.020404}
  {10.1103/physrevb.88.020404} (\bibinfo {year} {2013})\BibitemShut {NoStop}%
\bibitem [{\citenamefont {Rusz}\ \emph {et~al.}(2005)\citenamefont {Rusz},
  \citenamefont {Turek},\ and\ \citenamefont {Divi{\v{s}}}}]{Rusz2005}%
  \BibitemOpen
  \bibfield  {author} {\bibinfo {author} {\bibfnamefont {J.}~\bibnamefont
  {Rusz}}, \bibinfo {author} {\bibfnamefont {I.}~\bibnamefont {Turek}},\ and\
  \bibinfo {author} {\bibfnamefont {M.}~\bibnamefont {Divi{\v{s}}}},\
  }\bibfield  {journal} {\bibinfo  {journal} {Physical Review B}\ }\textbf
  {\bibinfo {volume} {71}},\ \href {https://doi.org/10.1103/physrevb.71.174408}
  {10.1103/physrevb.71.174408} (\bibinfo {year} {2005})\BibitemShut {NoStop}%
\bibitem [{\citenamefont {Rusz}\ \emph {et~al.}(2006)\citenamefont {Rusz},
  \citenamefont {Bergqvist}, \citenamefont {Kudrnovsky},\ and\ \citenamefont
  {Turek}}]{Rusz2006}%
  \BibitemOpen
  \bibfield  {author} {\bibinfo {author} {\bibfnamefont {J.}~\bibnamefont
  {Rusz}}, \bibinfo {author} {\bibfnamefont {L.}~\bibnamefont {Bergqvist}},
  \bibinfo {author} {\bibfnamefont {J.}~\bibnamefont {Kudrnovsky}},\ and\
  \bibinfo {author} {\bibfnamefont {I.}~\bibnamefont {Turek}},\ }\bibfield
  {journal} {\bibinfo  {journal} {Physical Review B}\ }\textbf {\bibinfo
  {volume} {73}},\ \href
  {https://doi.org/https://doi.org/10.1103/PhysRevB.73.214412}
  {https://doi.org/10.1103/PhysRevB.73.214412} (\bibinfo {year}
  {2006})\BibitemShut {NoStop}%
\bibitem [{\citenamefont {Liechtenstein}\ \emph {et~al.}(1987)\citenamefont
  {Liechtenstein}, \citenamefont {Katsnelson}, \citenamefont {Antropov},\ and\
  \citenamefont {Gubanov}}]{Liechtenstein1987}%
  \BibitemOpen
  \bibfield  {author} {\bibinfo {author} {\bibfnamefont {A.}~\bibnamefont
  {Liechtenstein}}, \bibinfo {author} {\bibfnamefont {M.}~\bibnamefont
  {Katsnelson}}, \bibinfo {author} {\bibfnamefont {V.}~\bibnamefont
  {Antropov}},\ and\ \bibinfo {author} {\bibfnamefont {V.}~\bibnamefont
  {Gubanov}},\ }\href {https://doi.org/10.1016/0304-8853(87)90721-9} {\bibfield
   {journal} {\bibinfo  {journal} {Journal of Magnetism and Magnetic
  Materials}\ }\textbf {\bibinfo {volume} {67}},\ \bibinfo {pages} {65}
  (\bibinfo {year} {1987})}\BibitemShut {NoStop}%
\bibitem [{\citenamefont {Solovyev}(2021)}]{Solovyev2021}%
  \BibitemOpen
  \bibfield  {author} {\bibinfo {author} {\bibfnamefont {I.~V.}\ \bibnamefont
  {Solovyev}},\ }\href {https://doi.org/10.1103/physrevb.103.104428} {\bibfield
   {journal} {\bibinfo  {journal} {Physical Review B}\ }\textbf {\bibinfo
  {volume} {103}},\ \bibinfo {pages} {104428} (\bibinfo {year}
  {2021})}\BibitemShut {NoStop}%
\bibitem [{\citenamefont {Nolting}\ and\ \citenamefont
  {Ramakanth}(2009)}]{Nolting2009}%
  \BibitemOpen
  \bibfield  {author} {\bibinfo {author} {\bibfnamefont {W.}~\bibnamefont
  {Nolting}}\ and\ \bibinfo {author} {\bibfnamefont {A.}~\bibnamefont
  {Ramakanth}},\ }\href@noop {} {\emph {\bibinfo {title} {Quantum Theory of
  Magnetism}}}\ (\bibinfo  {publisher} {Springer},\ \bibinfo {year}
  {2009})\BibitemShut {NoStop}%
\bibitem [{\citenamefont {Yonezawa}(1968)}]{Yonezawa1968}%
  \BibitemOpen
  \bibfield  {author} {\bibinfo {author} {\bibfnamefont {F.}~\bibnamefont
  {Yonezawa}},\ }\href {https://doi.org/10.1143/ptp.40.734} {\bibfield
  {journal} {\bibinfo  {journal} {Progress of Theoretical Physics}\ }\textbf
  {\bibinfo {volume} {40}},\ \bibinfo {pages} {734} (\bibinfo {year}
  {1968})}\BibitemShut {NoStop}%
\bibitem [{\citenamefont {Matsubara}(1973)}]{Matsubara1973}%
  \BibitemOpen
  \bibfield  {author} {\bibinfo {author} {\bibfnamefont {T.}~\bibnamefont
  {Matsubara}},\ }\href {https://doi.org/10.1143/ptps.53.202} {\bibfield
  {journal} {\bibinfo  {journal} {Progress of Theoretical Physics Supplement}\
  }\textbf {\bibinfo {volume} {53}},\ \bibinfo {pages} {202} (\bibinfo {year}
  {1973})}\BibitemShut {NoStop}%
\bibitem [{\citenamefont {Tang}\ and\ \citenamefont
  {Nolting}(2006)}]{Tang2006}%
  \BibitemOpen
  \bibfield  {author} {\bibinfo {author} {\bibfnamefont {G.~X.}\ \bibnamefont
  {Tang}}\ and\ \bibinfo {author} {\bibfnamefont {W.}~\bibnamefont {Nolting}},\
  }\bibfield  {journal} {\bibinfo  {journal} {Physical Review B}\ }\textbf
  {\bibinfo {volume} {73}},\ \href {https://doi.org/10.1103/physrevb.73.024415}
  {10.1103/physrevb.73.024415} (\bibinfo {year} {2006})\BibitemShut {NoStop}%
\bibitem [{\citenamefont {Bouzerar}\ and\ \citenamefont
  {Bruno}(2002)}]{Bouzerar2002}%
  \BibitemOpen
  \bibfield  {author} {\bibinfo {author} {\bibfnamefont {G.}~\bibnamefont
  {Bouzerar}}\ and\ \bibinfo {author} {\bibfnamefont {P.}~\bibnamefont
  {Bruno}},\ }\bibfield  {journal} {\bibinfo  {journal} {Physical Review B}\
  }\textbf {\bibinfo {volume} {66}},\ \href
  {https://doi.org/10.1103/physrevb.66.014410} {10.1103/physrevb.66.014410}
  (\bibinfo {year} {2002})\BibitemShut {NoStop}%
\bibitem [{\citenamefont {Blöchl}\ \emph {et~al.}(1994)\citenamefont
  {Blöchl}, \citenamefont {Jepsen},\ and\ \citenamefont
  {Andersen}}]{Bloechl1994}%
  \BibitemOpen
  \bibfield  {author} {\bibinfo {author} {\bibfnamefont {P.~E.}\ \bibnamefont
  {Blöchl}}, \bibinfo {author} {\bibfnamefont {O.}~\bibnamefont {Jepsen}},\
  and\ \bibinfo {author} {\bibfnamefont {O.~K.}\ \bibnamefont {Andersen}},\
  }\href {https://doi.org/10.1103/physrevb.49.16223} {\bibfield  {journal}
  {\bibinfo  {journal} {Physical Review B}\ }\textbf {\bibinfo {volume} {49}},\
  \bibinfo {pages} {16223} (\bibinfo {year} {1994})}\BibitemShut {NoStop}%
\bibitem [{\citenamefont {Ger\v{s}gorin}(1931)}]{Gersgorin1931}%
  \BibitemOpen
  \bibfield  {author} {\bibinfo {author} {\bibfnamefont {S.}~\bibnamefont
  {Ger\v{s}gorin}},\ }\href@noop {} {\bibfield  {journal} {\bibinfo  {journal}
  {{Bulletin de l’Acad´emie des Sciences de l’URSS}}\ } (\bibinfo {year}
  {1931})}\BibitemShut {NoStop}%
\bibitem [{\citenamefont {Perdew}\ \emph {et~al.}(1996)\citenamefont {Perdew},
  \citenamefont {Burke},\ and\ \citenamefont {Ernzerhof}}]{Perdew1996}%
  \BibitemOpen
  \bibfield  {author} {\bibinfo {author} {\bibfnamefont {J.~P.}\ \bibnamefont
  {Perdew}}, \bibinfo {author} {\bibfnamefont {K.}~\bibnamefont {Burke}},\ and\
  \bibinfo {author} {\bibfnamefont {M.}~\bibnamefont {Ernzerhof}},\ }\href
  {https://doi.org/10.1103/physrevlett.77.3865} {\bibfield  {journal} {\bibinfo
   {journal} {Physical Review Letters}\ }\textbf {\bibinfo {volume} {77}},\
  \bibinfo {pages} {3865} (\bibinfo {year} {1996})}\BibitemShut {NoStop}%
\bibitem [{\citenamefont {Lüders}\ \emph {et~al.}(2001)\citenamefont
  {Lüders}, \citenamefont {Ernst}, \citenamefont {Temmerman}, \citenamefont
  {Szotek},\ and\ \citenamefont {Durham}}]{Luders2001}%
  \BibitemOpen
  \bibfield  {author} {\bibinfo {author} {\bibfnamefont {M.}~\bibnamefont
  {Lüders}}, \bibinfo {author} {\bibfnamefont {A.}~\bibnamefont {Ernst}},
  \bibinfo {author} {\bibfnamefont {W.~M.}\ \bibnamefont {Temmerman}}, \bibinfo
  {author} {\bibfnamefont {Z.}~\bibnamefont {Szotek}},\ and\ \bibinfo {author}
  {\bibfnamefont {P.~J.}\ \bibnamefont {Durham}},\ }\href
  {https://doi.org/10.1088/0953-8984/13/38/305} {\bibfield  {journal} {\bibinfo
   {journal} {Journal of Physics: Condensed Matter}\ }\textbf {\bibinfo
  {volume} {13}},\ \bibinfo {pages} {8587} (\bibinfo {year}
  {2001})}\BibitemShut {NoStop}%
\bibitem [{\citenamefont {Geilhufe}\ \emph {et~al.}(2015)\citenamefont
  {Geilhufe}, \citenamefont {Achilles}, \citenamefont {Köbis}, \citenamefont
  {Arnold}, \citenamefont {Mertig}, \citenamefont {Hergert},\ and\
  \citenamefont {Ernst}}]{Geilhufe2015}%
  \BibitemOpen
  \bibfield  {author} {\bibinfo {author} {\bibfnamefont {M.}~\bibnamefont
  {Geilhufe}}, \bibinfo {author} {\bibfnamefont {S.}~\bibnamefont {Achilles}},
  \bibinfo {author} {\bibfnamefont {M.~A.}\ \bibnamefont {Köbis}}, \bibinfo
  {author} {\bibfnamefont {M.}~\bibnamefont {Arnold}}, \bibinfo {author}
  {\bibfnamefont {I.}~\bibnamefont {Mertig}}, \bibinfo {author} {\bibfnamefont
  {W.}~\bibnamefont {Hergert}},\ and\ \bibinfo {author} {\bibfnamefont
  {A.}~\bibnamefont {Ernst}},\ }\href
  {https://doi.org/10.1088/0953-8984/27/43/435202} {\bibfield  {journal}
  {\bibinfo  {journal} {Journal of Physics: Condensed Matter}\ }\textbf
  {\bibinfo {volume} {27}},\ \bibinfo {pages} {435202} (\bibinfo {year}
  {2015})}\BibitemShut {NoStop}%
\bibitem [{\citenamefont {Hoffmann}\ \emph {et~al.}(2020)\citenamefont
  {Hoffmann}, \citenamefont {Ernst}, \citenamefont {Hergert}, \citenamefont
  {Antonov}, \citenamefont {Adeagbo}, \citenamefont {Geilhufe},\ and\
  \citenamefont {Hamed}}]{Hoffmann2020}%
  \BibitemOpen
  \bibfield  {author} {\bibinfo {author} {\bibfnamefont {M.}~\bibnamefont
  {Hoffmann}}, \bibinfo {author} {\bibfnamefont {A.}~\bibnamefont {Ernst}},
  \bibinfo {author} {\bibfnamefont {W.}~\bibnamefont {Hergert}}, \bibinfo
  {author} {\bibfnamefont {V.~N.}\ \bibnamefont {Antonov}}, \bibinfo {author}
  {\bibfnamefont {W.~A.}\ \bibnamefont {Adeagbo}}, \bibinfo {author}
  {\bibfnamefont {R.~M.}\ \bibnamefont {Geilhufe}},\ and\ \bibinfo {author}
  {\bibfnamefont {H.~B.}\ \bibnamefont {Hamed}},\ }\href
  {https://doi.org/10.1002/pssb.201900671} {\bibfield  {journal} {\bibinfo
  {journal} {physica status solidi (b)}\ }\textbf {\bibinfo {volume} {257}},\
  \bibinfo {pages} {1900671} (\bibinfo {year} {2020})},\ \Eprint
  {https://arxiv.org/abs/https://onlinelibrary.wiley.com/doi/pdf/10.1002/pssb.201900671}
  {https://onlinelibrary.wiley.com/doi/pdf/10.1002/pssb.201900671} \BibitemShut
  {NoStop}%
\bibitem [{\citenamefont {Soven}(1967)}]{Soven1967}%
  \BibitemOpen
  \bibfield  {author} {\bibinfo {author} {\bibfnamefont {P.}~\bibnamefont
  {Soven}},\ }\href {https://doi.org/10.1103/physrev.156.809} {\bibfield
  {journal} {\bibinfo  {journal} {Physical Review}\ }\textbf {\bibinfo {volume}
  {156}},\ \bibinfo {pages} {809} (\bibinfo {year} {1967})}\BibitemShut
  {NoStop}%
\bibitem [{\citenamefont {Gyorffy}(1972)}]{Gyorffy1972}%
  \BibitemOpen
  \bibfield  {author} {\bibinfo {author} {\bibfnamefont {B.~L.}\ \bibnamefont
  {Gyorffy}},\ }\href {https://doi.org/10.1103/physrevb.5.2382} {\bibfield
  {journal} {\bibinfo  {journal} {Physical Review B}\ }\textbf {\bibinfo
  {volume} {5}},\ \bibinfo {pages} {2382} (\bibinfo {year} {1972})}\BibitemShut
  {NoStop}%
\bibitem [{\citenamefont {Turek}\ \emph {et~al.}(2006)\citenamefont {Turek},
  \citenamefont {Kudrnovsk{\'{y}}}, \citenamefont {Drchal},\ and\ \citenamefont
  {Bruno}}]{Turek2006}%
  \BibitemOpen
  \bibfield  {author} {\bibinfo {author} {\bibfnamefont {I.}~\bibnamefont
  {Turek}}, \bibinfo {author} {\bibfnamefont {J.}~\bibnamefont
  {Kudrnovsk{\'{y}}}}, \bibinfo {author} {\bibfnamefont {V.}~\bibnamefont
  {Drchal}},\ and\ \bibinfo {author} {\bibfnamefont {P.}~\bibnamefont
  {Bruno}},\ }\href {https://doi.org/10.1080/14786430500504048} {\bibfield
  {journal} {\bibinfo  {journal} {Philosophical Magazine}\ }\textbf {\bibinfo
  {volume} {86}},\ \bibinfo {pages} {1713} (\bibinfo {year}
  {2006})}\BibitemShut {NoStop}%
\bibitem [{\citenamefont {Kokorina}\ and\ \citenamefont
  {Medvedev}(2013)}]{Kokorina2013}%
  \BibitemOpen
  \bibfield  {author} {\bibinfo {author} {\bibfnamefont {E.}~\bibnamefont
  {Kokorina}}\ and\ \bibinfo {author} {\bibfnamefont {M.}~\bibnamefont
  {Medvedev}},\ }\href {https://doi.org/10.1016/j.physb.2013.01.035} {\bibfield
   {journal} {\bibinfo  {journal} {Physica B: Condensed Matter}\ }\textbf
  {\bibinfo {volume} {416}},\ \bibinfo {pages} {29} (\bibinfo {year}
  {2013})}\BibitemShut {NoStop}%
\bibitem [{\citenamefont {Essenberger}\ \emph {et~al.}(2012)\citenamefont
  {Essenberger}, \citenamefont {Buczek}, \citenamefont {Ernst}, \citenamefont
  {Sandratskii},\ and\ \citenamefont {Gross}}]{Essenberger2012}%
  \BibitemOpen
  \bibfield  {author} {\bibinfo {author} {\bibfnamefont {F.}~\bibnamefont
  {Essenberger}}, \bibinfo {author} {\bibfnamefont {P.}~\bibnamefont {Buczek}},
  \bibinfo {author} {\bibfnamefont {A.}~\bibnamefont {Ernst}}, \bibinfo
  {author} {\bibfnamefont {L.}~\bibnamefont {Sandratskii}},\ and\ \bibinfo
  {author} {\bibfnamefont {E.~K.~U.}\ \bibnamefont {Gross}},\ }\bibfield
  {journal} {\bibinfo  {journal} {Physical Review B}\ }\textbf {\bibinfo
  {volume} {86}},\ \href {https://doi.org/10.1103/physrevb.86.060412}
  {10.1103/physrevb.86.060412} (\bibinfo {year} {2012})\BibitemShut {NoStop}%
\bibitem [{\citenamefont {Hüller}(1986)}]{Hueller1986}%
  \BibitemOpen
  \bibfield  {author} {\bibinfo {author} {\bibfnamefont {K.}~\bibnamefont
  {Hüller}},\ }\href {https://doi.org/10.1016/0304-8853(86)90048-x} {\bibfield
   {journal} {\bibinfo  {journal} {Journal of Magnetism and Magnetic
  Materials}\ }\textbf {\bibinfo {volume} {61}},\ \bibinfo {pages} {347}
  (\bibinfo {year} {1986})}\BibitemShut {NoStop}%
\bibitem [{\citenamefont {Knoll}\ \emph {et~al.}(1990)\citenamefont {Knoll},
  \citenamefont {Thomsen}, \citenamefont {Cardona},\ and\ \citenamefont
  {Murugaraj}}]{Knoll1990}%
  \BibitemOpen
  \bibfield  {author} {\bibinfo {author} {\bibfnamefont {P.}~\bibnamefont
  {Knoll}}, \bibinfo {author} {\bibfnamefont {C.}~\bibnamefont {Thomsen}},
  \bibinfo {author} {\bibfnamefont {M.}~\bibnamefont {Cardona}},\ and\ \bibinfo
  {author} {\bibfnamefont {P.}~\bibnamefont {Murugaraj}},\ }\href
  {https://doi.org/10.1103/physrevb.42.4842} {\bibfield  {journal} {\bibinfo
  {journal} {Physical Review B}\ }\textbf {\bibinfo {volume} {42}},\ \bibinfo
  {pages} {4842} (\bibinfo {year} {1990})}\BibitemShut {NoStop}%
\end{thebibliography}%

\end{document}